%% file: main.tex
\documentclass[lettersize,journal]{IEEEtran}
\usepackage{amsmath,amsfonts}
\usepackage{algorithmic}
\usepackage{algorithm}
\usepackage{array}
\usepackage[caption=false,font=normalsize,labelfont=sf,textfont=sf]{subfig}
\usepackage{textcomp}
\usepackage{stfloats}
\usepackage{url}
\usepackage{verbatim}
\usepackage{graphicx}
\usepackage{tikz}
\newcommand{\circled}[2][10pt]{%
  \tikz[baseline=(char.base)]\node[draw=black,circle,inner sep=#1, fill=black, text=white] (char) {#2};%
}
\usepackage{enumitem}
\usepackage{cite}
\usepackage{xcolor}

\definecolor{comment}{RGB}{251, 236, 93}

\usepackage{array}
\usepackage{colortbl}
\newcolumntype{C}[1]{>{\centering\arraybackslash}m{#1}}

\begin{document}

\title{FlexSpIM: An Event-Based Digital Compute-In-Memory Accelerator\\
with Flexible Operand Resolution and Layer-Wise Hybrid Stationarity}

\author{\IEEEauthorblockN{Nicolas Chauvaux, \textit{Graduate Student Member, IEEE}, Adrian Kneip, \textit{Member, IEEE}, and Charlotte Frenkel, \textit{Senior Member, IEEE}}
\vspace*{-0.5cm}

\thanks{Manuscript received Month xx, 2026; revised Month xx, 2026; accepted Month xx, 2026. This work was co-funded by Prophesee and by the Dutch government as an HTSM-TKI project, and supported by DelftBlue and Cloud4Research to get access to the computational resources required to obtain the system-level results.
This article was recommended by Associate Editor [Surname Lastname]. \textit{(Corresponding author: Nicolas Chauvaux.)}

Nicolas Chauvaux and Charlotte Frenkel are with the Department of Microelectronics, Delft University of Technology, 2628 CD Delft, The Netherlands (e-mail:
n.chauvaux@tudelft.nl).\\
Adrian Kneip is with the Department of Microelectronics, Delft University of Technology, 2628CD Delft, The Netherlands, and with the Department of Electrical Engineering, KU Leuven, 3001 Leuven, Belgium.}
}

\maketitle

\input{text/abstract}
\input{text/introduction}
\input{text/background}
\input{text/macro_level}
\input{text/system_level}
\input{text/results}
\input{text/conclusion}
\input{text/acknowledgments}

\vspace{-0.25cm}
\bibliographystyle{IEEEtran}
\bibliography{references_wo_abbreviation}

\input{text/biography}
\end{document}

%% file: text/abstract.tex
\begin{abstract}
Compute-in-memory (CIM) accelerators for spiking neural networks (SNNs) offer a promising solution for achieving $\mu$s-level inference latency and ultra-low energy in edge vision applications. However, their limited flexibility at both circuit and system levels restricts their deployment across diverse workloads. This work introduces FlexSpIM, a digital CIM architecture supporting arbitrary operand resolution and shape within a unified storage for weights and neuron states (i.e., membrane potentials). These circuit-level capabilities enable a layer-level hybrid weight- and output-stationary dataflow, maximizing operand reuse and reducing costly on- and off-chip data movement during SNN execution. Measurement results from a fabricated FlexSpIM prototype in 40-nm CMOS demonstrate competitive 1-bit-normalized energy efficiency and higher throughput compared with prior fixed-precision digital CIM-based SNN accelerators, while providing bitwise resolution reconfiguration. Evaluated on the IBM DVS gesture dataset, FlexSpIM achieves 95.8\% accuracy while enabling up to 45\% energy and 52\% latency reductions in large-scale systems compared with fixed stationarity approaches.
\end{abstract}

\begin{IEEEkeywords}
Digital compute-in-memory, spiking neural networks, flexible operand resolution, hybrid-stationary dataflow.
\end{IEEEkeywords}
\vspace{-0.25cm}

%% file: text/introduction.tex
\section{Introduction}
\begin{figure}[t]
    \centering
    \includegraphics[width=\linewidth]{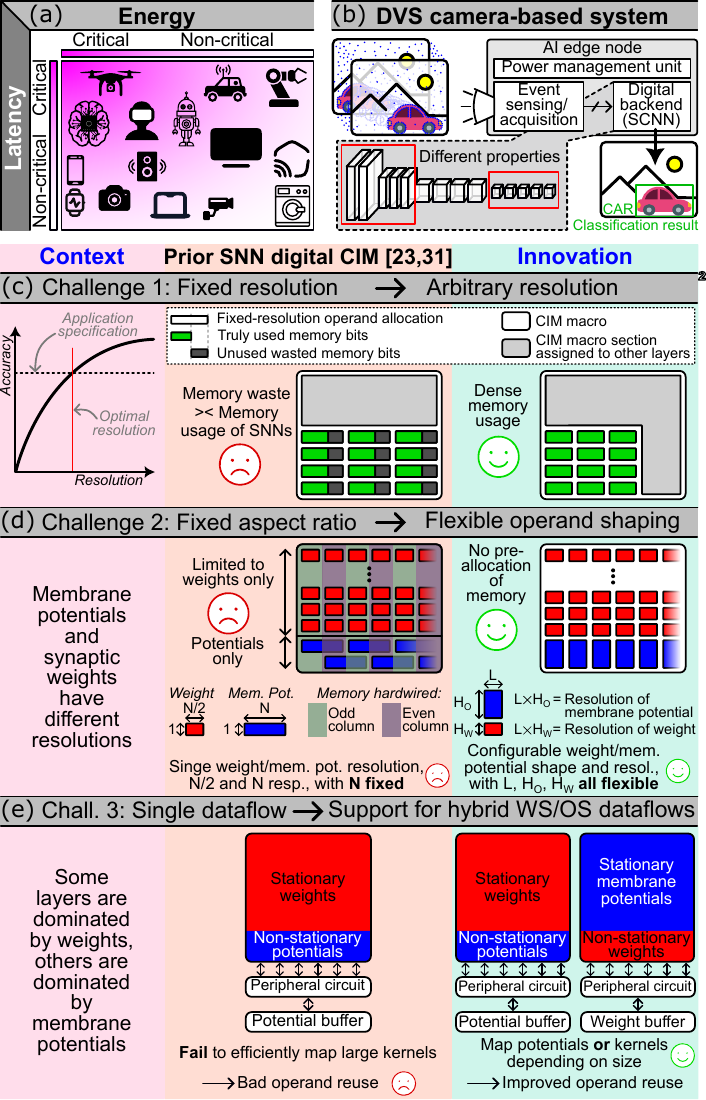}
    \vspace{-0.4cm}
    \caption{(a) Application space of artificial neural networks, categorized according to latency and energy requirements. (b) Simplified edge vision system using an event-based camera as input and a spiking convolutional neural network (SCNN) backend for object tracking, highlighting the diversity of layer shapes encountered during execution. (c) Three key challenges addressed in this work for efficient execution of SCNNs on digital compute-in-memory (CIM) architectures.}
    \label{fig:intro}
    \vspace{-0.7cm}
\end{figure}

\IEEEPARstart{A}{rtificial} neural networks (ANNs) have become ubiquitous across a wide range of applications. Taking the example of image classification, their application can be categorized based on the criticality of their energy and latency requirements, as illustrated in Fig.~\ref{fig:intro}(a). Compact models optimized for mobile CPUs, such as MobileNetV3, already require hundreds of millions of multiply-accumulate (MAC) operations and millions of parameters~\cite{MobileNetv3}. As transferring these parameters can account for up to $60\%$ of the total power of modern workloads~\cite{ZigZag}, recent works started exploiting sparsity to reduce energy consumption and latency~\cite{DPIM, ONYX}.

Exploiting sparsity is achieved by bypassing \textit{zero} operations and/or using compression techniques to minimize memory requirements~\cite{SSCM}. Several works exploit input~\cite{Eyeriss, Cnvlutin} or weight~\cite{weight-sparsity-CNN} unstructured sparsity, while others combine both~\cite{input-weight-sparsity-CNN, Eyeriss-v2, Cnvlutin2}, at the expense of increasing datapath and memory controller complexity. State-of-the-art sparsity exploitation techniques have demonstrated that sparsity levels of up to $70\%$ for inputs and $80\%$ for weights can be exploited without degrading accuracy on the ImageNet dataset~\cite{accuracy-loss-with-sparsity}, which is shown by Parashar~\textit{et~al.} to translate to $2.3\times$ energy and $2.7\times$ latency improvements~\cite{input-weight-sparsity-CNN}. However, while sparsity levels above $90\%$ outline additional order-of-magnitude energy and latency savings, exploiting them at iso-accuracy remains an open challenge~\cite{accuracy-loss-with-sparsity}.

In order to exploit such extreme sparsity regimes, recent approaches shift from frame- to event-based vision with dynamic-vision-sensor (DVS) cameras (Fig.~\ref{fig:intro}(b)), in which individual pixels generate events only when their temporal contrast delta crosses a given threshold~\cite{DVS_CAMERA}. The \textit{active-pixel ratio}, i.e., the fraction of pixels generating at least one event within a given time window, typically ranges from below $1\%$ to around $10\%$, depending on scene dynamics, for a static and a moving camera respectively, in an autonomous driving scenario~\cite{event_active_pixels}. Combined with sub-100-$\mu$s end-to-end event generation latency~\cite{Prophesee}, this makes DVS cameras well suited for sparse and latency-critical vision applications, motivating for novel event-driven vision processing solutions.

Spiking convolutional neural networks (SCNNs) are gaining increasing interest for event-driven vision processing, and the efficient processing of \textit{spatiotemporal} information, thanks to the use of a stateful neuron with a membrane potential~\cite{BOTTOM_UP}. However, this comes at the expense of additional memory footprint as, in an SCNN, the neurons' membrane potentials need to be preserved over time, as opposed to conventional frame-based CNNs whose partial sums can be discarded after they have been used to update the downstream neurons. Indeed, assuming 8-bit weights and 16-bit membrane potentials, the memory overhead of SCNNs can amount to $25-67\%$ of the total memory footprint of the equivalent non-spiking CNN~\cite{VGG16, ALEXNET, MOBILENET}. Since memory accesses are typically the main bottleneck in terms of energy and latency, the membrane potential memory overhead of SCNNs is currently hindering claims related to their energy/latency advantage~\cite{SpQuant_SNN, LIF_overhead, SNN_efficiency}.
As keeping all weights and membrane potentials on-chip is not a practical solution for large SCNNs, it is key to find hardware architectures that minimize data movement \textit{at the system level}, taking into account external memory accesses (EMAs).

To that end, compute-in-memory (CIM) hardware for SCNNs has recently been proposed \cite{IMPULSE, ISSCC24, NEUROCIM, SPIKECIM, TETC23, DS_CIM, SNNIM, SRIF, SPIDR, 1T1C-EDRAM}, but their limited reconfigurability makes them fundamentally unsuited to the large diversity of SCNN layer specifications (Fig.~\ref{fig:intro}(b)). This either implies sub-optimal workload mapping to CIM hardware, leading to latency and energy penalties, or increased time-to-market due to the need for application-specific CIM hardware.
In this work, we propose FlexSpIM, a digital CIM-based accelerator for SNN inference with high flexibility at the circuit and system levels. It solves three core challenges compared to prior CIM-based SNNs:
\vspace*{-1mm}
\begin{enumerate}[leftmargin=*]
    \item while previous works only support a fixed resolution or a few pre-defined options, FlexSpIM supports a fully reconfigurable resolution for weights and membrane potentials, thereby expanding the exploration of the trade-off landscape between accuracy, energy efficiency, and memory footprint for SNN workloads~(Fig.~\ref{fig:intro}(c));
    \item while operand mapping is usually restricted to either fully bit-serial row-wise or fully bit-parallel column-wise, FlexSpIM allows for reconfigurable operand shapes to support different, non-proportional resolution values for weights and membrane potentials, which were otherwise constrained to fixed ratios~\cite{IMPULSE}~(Fig.~\ref{fig:intro}(d));
    \item while maximizing operand stationarity is key to reduce EMAs, prior CIM-based SNNs only support weight stationarity, and thus are ill-suited to layers that are bottlenecked by membrane-potential data movement (e.g., first layers of ResNet \cite{RESNET}). FlexSpIM introduces a unified weight/membrane potential memory that allows for a hybrid dataflow, making the best out of both weight stationarity (WS) and output (i.e., membrane potential) stationarity (OS) on a per-layer basis. This minimizes operand replacement in the CIM macro for the selected workload, thereby directly alleviating the data movement efficiency bottleneck of previous work~(Fig.~\ref{fig:intro}(d)).
\end{enumerate}

\noindent The remainder of this paper is organized as follows. Section~\ref{sec:background} reviews the background of spiking neurons for SCNNs and CIM architectures. Section~\ref{sec:macro} presents the proposed FlexSpIM architecture and discusses how it addresses the key challenges identified above. Section~\ref{sec:system} evaluates the proposed approach at the system level and compares it with a conventional weight-stationary dataflow. Finally, Section~\ref{sec:conclusion} concludes the paper.

%% file: text/background.tex
\section{Background}\label{sec:background}
Understanding the challenges of applying CIM to SNNs first requires an overview of the underlying concepts. Section~\ref{sec:spiking_neural_network} introduces the fundamentals of SNNs, and discusses the main execution methodologies and their tradeoffs. Then, Section Section~\ref{sec:CIM_for_SNN} presents the two principal classes of CIM architectures and their suitability for SNN acceleration.

\subsection{Spiking Neural Networks (SNNs)}
\label{sec:spiking_neural_network}
\subsubsection{Spiking neurons}\label{sec:spiking_neurons}
SNNs aim to mimic biological processing through spike-based events, introducing two key features. On the one hand, spiking neurons are stateful: a membrane potential variable is maintained across time for each neuron. On the other hand, activation functions are spiking: an input spike increments the neuron's membrane potential by the associated synaptic weight $w$ thereby relying on accumulate instead of multiply-and-accumulate operations. Neurons only fire an output spike when their potential reaches a predetermined threshold, after which it is reset (Fig.~\ref{fig:snn_background}(a)).
In this work, we consider a standard \textit{integrate-and-fire} (I\&F) neuron model \cite{SNN_MIT}. The evolution of the membrane potential $V^{j}_{\mathrm{mem}}$ of neuron $j$ is described as

\begin{equation}
{\frac{\mathrm{dV}^{j}_{\mathrm{mem}}(t)}{\mathrm{dt}}} =  \sum_{i=1}^{N}w_{i,j} \sum_{o \in O^{i}(t)} \delta(t-t_{o}^{i}),
\label{eq:neuron_integration}
\end{equation}

\noindent where $N$ is the number of pre-synaptic neurons, $w_{i,j}$ are synaptic weights, and $O^{i}(t)$ denotes the spike train from pre-synaptic neuron $i$, with spike times $t_o^i$ for each spike $o\in O^{i}$. The firing condition of neuron $j$ is described as

\begin{equation}
{O^{j}(t) = {\theta}(V^{j}_{\mathrm{mem}}(t))} = \begin{cases}
1&{\text{if}}\ V^{j}_{\mathrm{mem}}(t) \geq V_{\mathrm{th}}, \\ 
{0}&{\text{otherwise,}} 
\end{cases}
\label{eq:neuron_threshold}
\end{equation}

\noindent where $V_{\mathrm{th}}$ is the neuron firing threshold, and $\theta(\cdot)$ is the Heaviside function that handles the output spike generation and the reset of the membrane potential. Typical reset mechanisms are either a reset to zero, or a reset by subtraction of the firing threshold $V_{\mathrm{th}}$. The latter was shown to increase accuracy~\cite{RMP-SNN} and will be adopted in this work. As the I\&F model relies only on additions and comparisons, it is a low-footprint model often used in custom hardware implementations~\cite{ANP-I, IMPULSE, SNPU, C-DNN}.

\subsubsection{SNN execution}\label{sec:snn_execution}
\begin{figure}[t]
    \centering
    \includegraphics[width=\linewidth]{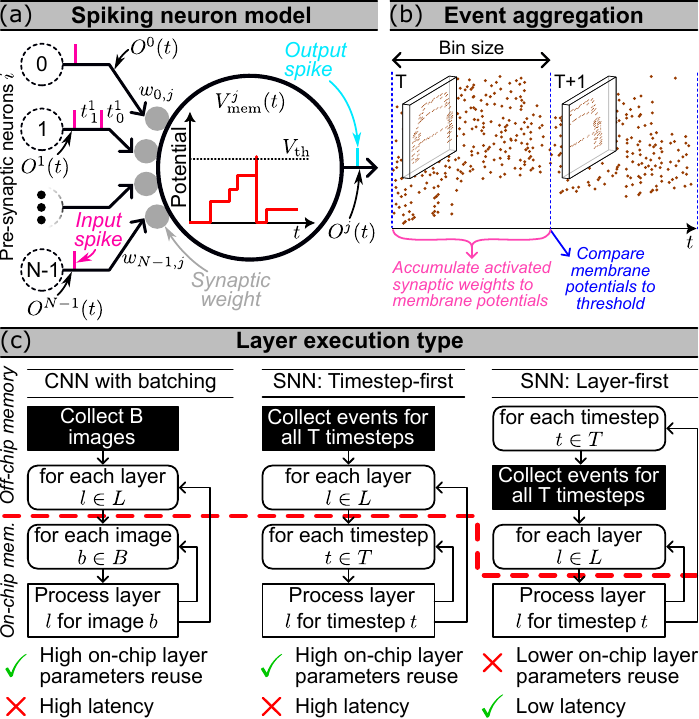}
    \caption{(a) Illustration of an integrate-and-fire (I\&F) neuron $j$ connected to $N$ pre-synaptic neurons indexed by $i$. (b) Illustration of the event aggregation process. (c) Overview of the main layer execution strategies: CNN-style batching for comparison (left), timestep-first execution (middle), and layer-first execution (right).}
    \label{fig:snn_background}
    \vspace{-0.3cm}
\end{figure}

SNN execution begins by aggregating events into bins (Fig.~\ref{fig:snn_background}(b))~\cite{timestep}, which segment the time dimension into timesteps. After each timestep, every neuron's membrane potential is compared against its threshold. A small bin size, down to a single event per bin for fully event-driven processing, increases this comparison overhead but reduces latency. Besides this bin size, SNN layer execution can follow two strategies (Fig.~\ref{fig:snn_background}(c)). In \textit{timestep-first} execution, analogous to CNN batching, multiple timesteps are executed per layer to maximize weight and membrane-potential reuse when not all layers fit on-chip, often the case due to limited on-chip memory~\cite{ISSCC24}. In \textit{layer-first} execution, all layers are instead executed for a single timestep before advancing to the next, offering lower latency at the cost of higher memory traffic. The former is inadequate for SNNs, since it loses the low-latency advantage we aim to exploit from DVS cameras. We therefore adopt layer-first execution, balancing latency and energy efficiency, which calls for an efficient dataflow such as the one proposed in this paper.

\subsection{CIM for SNNs}\label{sec:CIM_for_SNN}
\subsubsection{In-array vs. in-column CIM}\label{sec:digital_vs_analog}

\begin{figure}[t]
    \centering
    \includegraphics[width=\linewidth]{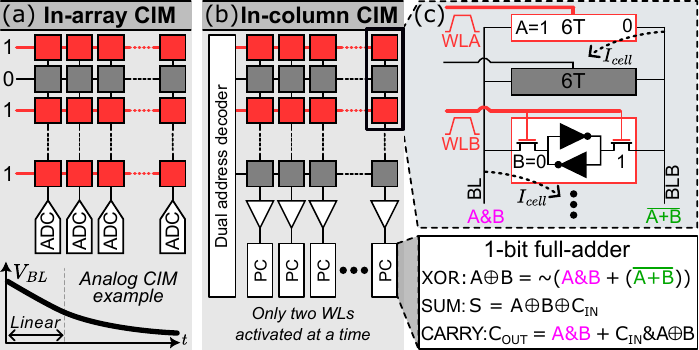}
    \caption{(a) In-array CIM principle, illustrated for analog CIM, with multi-WL activation and MVM representation as BL voltage. (b) In-column CIM principle with dual WL activation. (c) Impact of the activated bitcells on the bitlines and equations of a 1-bit full adder.}
    \label{fig:analog_digital_cim_explanation}
    \vspace{-0.1cm} 
\end{figure}

To reduce the latency and energy cost of data movement, compute-in-memory (CIM) approaches perform computation within memory arrays and/or their periphery by activating multiple wordlines (WLs) at once. Two orthogonal paradigms exist: CIM can either be in-array or in-column, whose distinction is particularly relevant for SNNs.

\textit{In-array CIM} (Fig.~\ref{fig:analog_digital_cim_explanation}(a)) performs matrix-vector multiplications (MVMs) massively in parallel directly in the memory array \cite{IMPACT}. Typically, analog CIMs activate multiple WLs in parallel, and the MVM result accumulates as charges or current on bitlines (BLs) before a final ADC conversion. Digital CIMs operate similarly, replacing the analog accumulation and ADCs by (approximate) multipliers and adder trees \cite{HierCIM, D6CIM, DATE25} or directly performing a full operation near each bitcell \cite{COLONNADE}.

\textit{In-column CIM} 
\begin{table}[t]
    \caption{Comparison between in-array and in-column CIM for CNN and SCNN operations.}
    \centering
    \includegraphics[width=\linewidth]{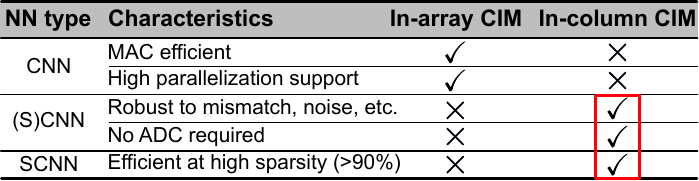}
    \label{tab:cim_comparison_table}
    \vspace{-0.5cm}
\end{table}
(Fig.~\ref{fig:analog_digital_cim_explanation}(b)) typically activates two WLs simultaneously, enabling digital logic operations between operands stored on the same BLs. In 6T SRAM, this produces operations such as $A\&B$ and $\overline{A|B}$, while peripheral circuits (PCs) extend this functionality to additions and multiplications, composing a full adder from $A\&B$ and $A|B$ without requiring $A$ and $B$ to be read out separately~\cite{NEURALCACHE, IMPULSE, VLSI26, DAC20, VLSI21}.

Table~\ref{tab:cim_comparison_table} compares both approaches for CNN and SCNN workloads. 
For CNNs, in-array CIM vastly outperforms in-column CIM thanks to its efficient acceleration of dense MVMs. Nonetheless, its computing accuracy can be limited by either analog impairments \cite{NONIDEALITIESCIM} or digital approximation techniques used to limit area overhead~\cite{approximate_adder, DIMC, Approximate}.
For SCNNs, the high sparsity of typical workloads (e.g., $>90\%$ in event-based vision sensors~\cite{event_active_pixels}) leads to underutilized in-array CIMs whose efficiency is capped by the ADC contribution \cite{ACIM_ZigZag}. In contrast, in-column CIMs inherently harness input sparsity, making them good candidates for sparse event-based processing. To abbreviate, we will thus refer to \textit{digital in-column CIM} as \textit{digital CIM} henceforth.

\subsubsection{State-of-the-art CIM for SNNs}\label{sec:soa_design}

\begin{figure}[t]
    \centering
    \includegraphics[width=\linewidth]{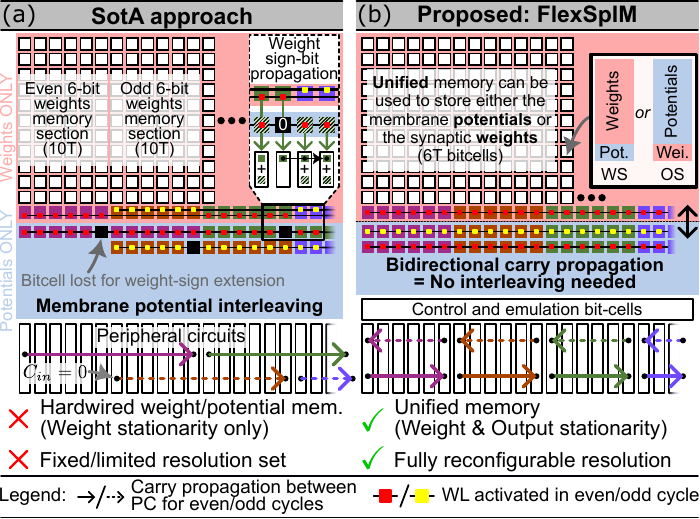}
    \caption{Operating principles of (a) state-of-the-art in-column CIM architectures for SNNs~\cite{IMPULSE,SPIDR} and (b) the proposed FlexSpIM architecture.}
    \label{fig:comparison_with_SoA}
    \vspace{-0.3cm}
\end{figure}
To the best of our knowledge, IMPULSE~\cite{IMPULSE}, shown in Fig.~\ref{fig:comparison_with_SoA}(a), is the only digital and in-column CIM accelerator specifically proposed for SNNs. Following the in-column CIM principle in Fig.~\ref{fig:analog_digital_cim_explanation}(b), synaptic weights and membrane potentials are stored in separate hardwired regions of the same memory array and added in-memory through carry propagation between PCs. To support 6-bit weights and 11-bit membrane potentials, IMPULSE interleaves odd/even potentials across two rows and maintains odd/even weights in a single row, processed in odd/even cycles, respectively. This requires separate WLs for odd and even weights and sacrifices one potential bit for weight-sign extraction. As a result, only a single set of weight/potential resolutions is supported and the memory organization is tied to a weight-stationary dataflow. SpiDR~\cite{SPIDR} extends IMPULSE to multiple macros and addresses its fixed-resolution limitation. However, it remains constrained to supporting only 4/7-bit, 6/11-bit, and 8/15-bit precision modes, as well as to maintaining separate hardwired regions of the CIM memory.

%% file: text/macro_level.tex
\section{Proposed Digital CIM Macro}\label{sec:macro}

\begin{figure*}
    \centering
    \includegraphics[width=\linewidth]{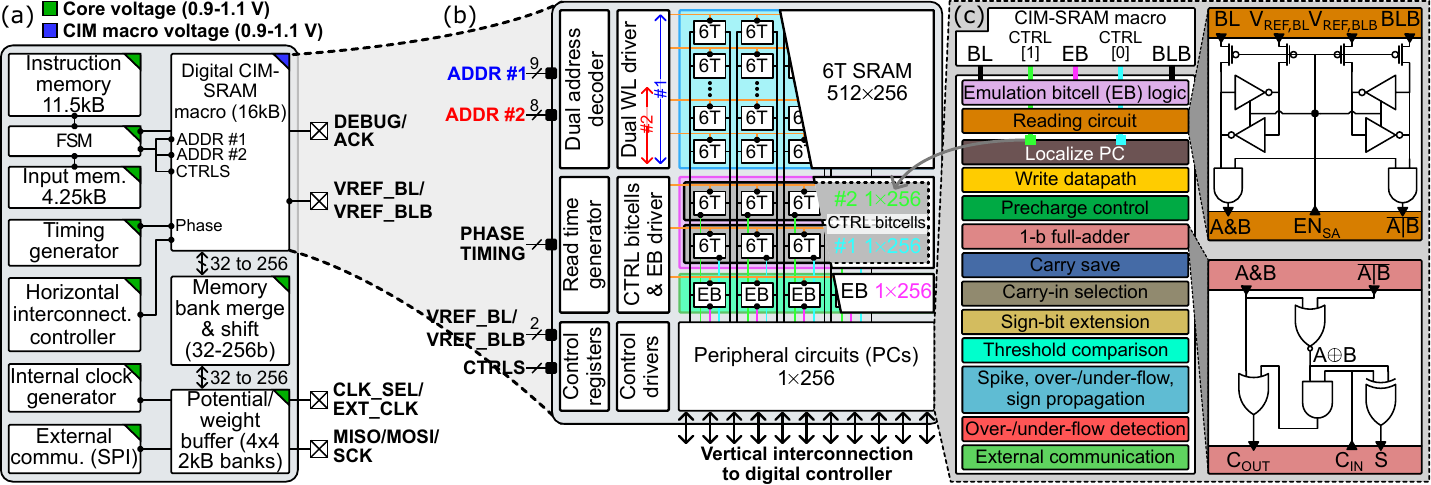}
    \caption{(a) FlexSpIM architecture. (b) Overview of the digital CIM macro architecture. (c) Module-level breakdown of the peripheral circuit, including detailed logic schematics of the read and 1-bit full-adder modules.}
    \label{fig:flexspim_architecture}
    \vspace{-0.3cm}
\end{figure*}

In this work, we propose the SRAM-based \textit{flexible spiking in-memory} (FlexSpIM) compute macro, which overcomes the limitations of fixed resolutions, mappings, and dataflows through a unified memory organization (Fig.~\ref{fig:comparison_with_SoA}(b)). 
We notably reshape membrane potentials into a $6\times2$ bitcell structure to enable bidirectional carry propagation, and thereby eliminate the membrane potential interleaving observed in IMPULSE/SpiDR. As a result, FlexSpIM is agnostic to the storage of weights or potentials, allowing it to support multiple dataflows, arbitrary operand resolutions through configurable PCs, and full membrane-potential precision. 

The top-level architecture of FlexSpIM is shown in Fig.~\ref{fig:flexspim_architecture}(a). It consists of a single $16$~kB digital CIM-SRAM macro, a $4.25$~kB input buffer for spikes, and a weight/potential buffer of $16$ banks of $2$~kB, which can be reshaped to match the CIM mapping pattern. A horizontal interconnect controller distributes shared neuron parameters to the CIM-SRAM, while a timing generator produces the phases of the CIM operation. The chip controller is based on a finite-state machine (FSM) that fetches instructions from instruction memory.

\vspace{-0.2cm}
\subsection{CIM-SRAM Overview}
The digital CIM-SRAM architecture (Fig.~\ref{fig:flexspim_architecture}(b)) combines a regular $512\times256$ 6T-SRAM bitcell array with per-column peripheral circuits (PCs), which are the core enabler of FlexSpIM. Each PC extends the bitwise AND/NOR logic performed by the in-column CIM-SRAM array into complex, precision-flexible operations, later detailed in Section~\ref{sec:operand_reshaping}. Three additional bitcells per column respectively store the 2-bit PC configuration (control bitcells) and the externally emulated 1-bit PC control (emulation bitcell, EB). A dual address decoder with full and half span generates WL controls for both regular read/write and dual-wordline in-memory operations. Finally, a dedicated in-SRAM controller handles the correct timing duration for WL activation.

\vspace{-0.3cm}
\subsection{CIM Operation}\label{sec:cim_operation}
Since input spikes directly activate the WLs to perform CIM operations, only weights and membrane potentials are stored in the CIM macro. To reduce data movement, one operand stays stationary while the other updates as needed. Unlike IMPULSE, which supports only weight-stationary (WS) execution (Section~\ref{sec:soa_design}), FlexSpIM also supports output-stationary (OS) execution by keeping membrane potentials stationary.

\begin{figure}
    \centering
    \includegraphics[width=\linewidth]{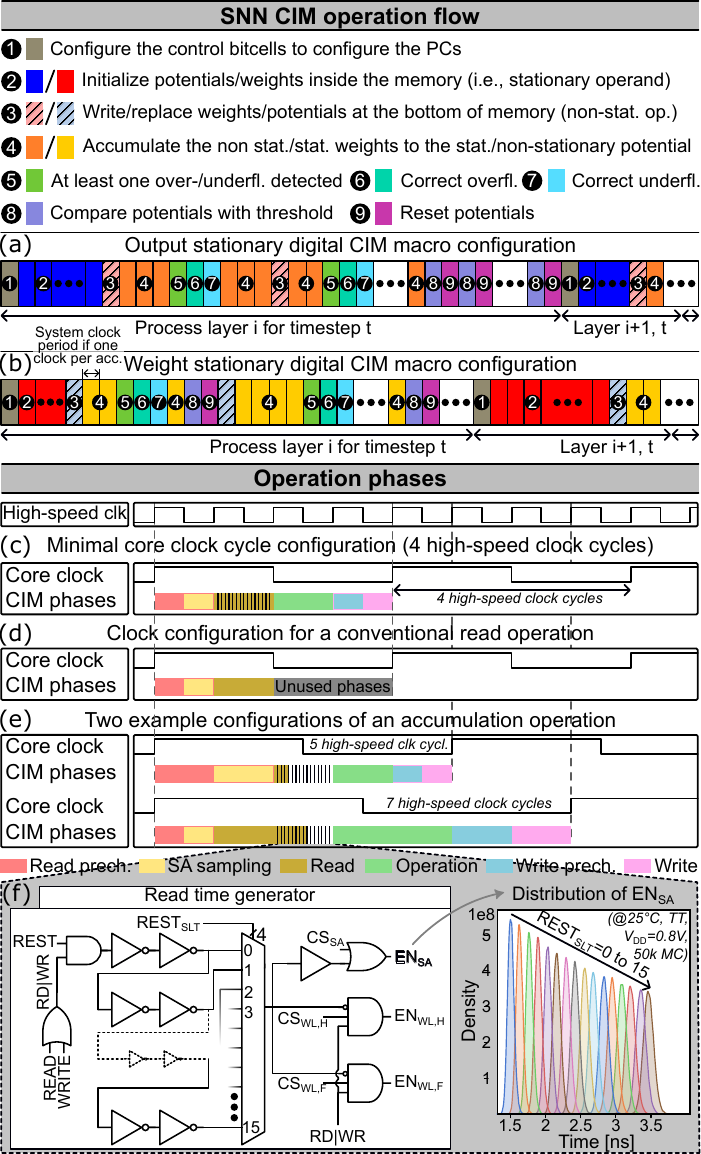}
    \caption{(a)-(b) SNN operation execution flow for the two supported dataflows. (c)-(e) Four phase timing configurations: (c) fastest accumulation, (d) conventional read, (e) two longer accumulation phase-timings. (f) Read-time generator schematic for fine-grained timing control of the read phase with post-layout Monte-Carlo simulation of all $15$ timing selections.} 
    \label{fig:cim_flow}
    \vspace{-0.8cm}
\end{figure}

The operation flow follows the layer-first schedule described in Section~\ref{sec:snn_execution}, and depends on the selected dataflow (Fig.~\ref{fig:cim_flow}(a)-(b)). Both OS and WS start by configuring the PC control bitcells for operand reshaping (Section~\ref{sec:configuration_setup}) and initialize the stationary operand (Section~\ref{sec:operand_initialization}). Non-stationary operands are stored at the bottom of the memory and updated during execution (Fig.~\ref{fig:comparison_with_SoA}(b), top right), with bits from only one non-stationary operand stored per memory column, maximizing the capacity available for stationary operands (Fig.~\ref{fig:comparison_with_SoA}(b), only two bits of the same membrane potential per column). CIM accumulation then proceeds by reusing the loaded non-stationary operand until replacement is required (Section~\ref{sec:accumulation}), with weight sign-bit extension and overflow/underflow correction applied when needed (Section~\ref{sec:bit_sign_extension} and Section~\ref{sec:overflow_underflow}, respectively). At each timestep, the membrane potential is thresholded and reset (Section~\ref{sec:comparison}). OS performs this after all events are processed, whereas WS performs it after the last accumulation event of each neuron to avoid reloading their membrane potential.

All operations share the same CIM macro and PC resources and are controlled through six fixed phases: \textit{BL precharge}, \textit{sense amplifier (SA) sampling}, \textit{WL(s) activation} (read/CIM), \textit{operation}, \textit{write precharge}, and \textit{write} (Fig.~\ref{fig:cim_flow}(c)-(e)). The duration of each phase is controlled by a finite-state machine (FSM) clocked by a programmable high-speed clock up to $1.2$~GHz. One cycle or more can be allocated to each phase, with a minimum of four cycles across all phases (Fig.~\ref{fig:cim_flow}(c)). The enabled phases depend on the operation type: Fig.~\ref{fig:cim_flow}(d) shows a conventional read operation, which does not require the write-related phases. Fig.~\ref{fig:cim_flow}(e) illustrates two of many possible timing variants. First, depending on the supply voltage, the precharge and sampling phases (and similarly the write-precharge and write phases) can share a single high-speed clock period or use distinct ones. Next, since WL activation is timing-critical for the read phase (see Section~\ref{sec:boolean_operation}), a configurable delay-chain-based read-time generator provides fine-grained control, validated through extensive post-layout Monte-Carlo simulations (Fig.~\ref{fig:cim_flow}(f)), with Fig.~\ref{fig:cim_flow}(e) showing two read-timing durations, the second spanning multiple clock cycles. Finally, to support various operand resolutions (see Section~\ref{sec:operand_reshaping}), the read/CIM phase time can also be adjusted to give the PCs appropriate time to complete the operation. The system's digital clock period is then defined by summing the durations of the six phases. The following sections detail each operation and its associated peripheral-circuit behavior, following Fig.~\ref{fig:cim_flow}(a)-(b).

\vspace{0.2cm}
\subsubsection{Configuration setup}\label{sec:configuration_setup}
Prior to operations of each layer, we configure the 2-bit control bitcells for the layer, operands and dataflow at hand, thereby defining the behavior of every PC during the following sequence of operations (Fig.~\ref{fig:cim_flow}(a)-(b)). Configuration data are supplied through the macro's I/O vertical interconnect (Fig.~\ref{fig:flexspim_architecture}(b)) and processed by the \textit{external-communication} and \textit{write-datapath} modules (Fig.~\ref{fig:flexspim_architecture}(c)). This operation uses only the write phases. Unlike regular memory accesses, the control bitcells are accessed through dedicated WLs rather than the dual address decoder.

\vspace{0.2cm}
\subsubsection{Operands initialization}\label{sec:operand_initialization}
Next, operands are written into the memory array using the same I/O interface. Stationary operands are mapped to the full range of addresses of the main decoder (i.e., via $\mathrm{ADDR \#1}$ in Fig.~\ref{fig:flexspim_architecture}), while non-stationary operands are stored at the bottom of the memory (i.e., via $\mathrm{ADDR \#2}$ in Fig.~\ref{fig:flexspim_architecture}). to facilitate multi-layer mapping (see Section~\ref{sec:system}), and overwrite non-stationary operands inside the CIM array. Stationary operands are written once and reused across all layers and timesteps, whereas non-stationary operands must be updated between layers and timesteps, and may require multiple updates within a single layer execution. Strategies for updating non-stationary operands to reduce latency and energy overheads are discussed in Section~\ref{sec:system}.

\vspace{0.2cm}
\subsubsection{Synaptic weight accumulation}\label{sec:accumulation}
Following these initialization steps, the accumulation operation takes place and is repeated, interleaved with non-stationary operand update if required (Fig.~\ref{fig:cim_flow}(a)), until all the events of the current timestep and layer are processed. The hardware operations required to process these events are described hereafter, from the basic Boolean CIM operations (Section~\ref{sec:boolean_operation}) and bit-serial accumulation (Section~\ref{sec:bit_serial}) to the proposed bit-parallel accumulation with operand reshaping (Section~\ref{sec:operand_reshaping}).

\paragraph{Boolean CIM operation}\label{sec:boolean_operation}
Unlike conventional reads, in-column CIM operations can create half-select write disturbances in 6T SRAM bitcells. If two activated bitcells on the same BL store different values, the cell storing 0 discharges the BL and may flip the cell storing 1 if the WL activation is too long. Limiting the activation time avoids bit flips but reduces the BL voltage swing. Therefore, latch-based sense amplifiers (SAs) with separate BL and BLB reference voltages are employed to ensure reliable Boolean result readout.

\begin{figure}
    \centering
    \includegraphics[width=\linewidth]{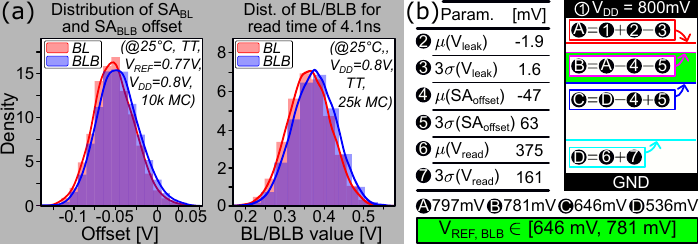}
    \caption{(a) Distribution of the BL and BLB sense amplifier offset from 10k Monte Carlo (MC) simulations, and distribution of the BL/BLB voltages after 4.1 ns of WL activation from 25k MC simulations. (b) Methodology to evaluate the margins of the reference levels employed by the read module.} 
    \label{fig:MC_simulations}
    \vspace{-0.2cm}
\end{figure}

To guarantee correct operation in the presence of local process variations, we extract the worst-case range of accepted reference voltages $V_{\mathrm{ref,BL/BLB}}$ with post-layout Monte-Carlo read-stability simulations (Fig.~\ref{fig:MC_simulations}(a)). 
For the BLB SA, the lower bound is derived from the maximum WL activation time without bit flips under a 3-$\sigma$ margin, resulting in $4.1$~ns at $V_\mathrm{DD} = 0.8$~V (Fig.~\ref{fig:MC_simulations}(b)). The corresponding $3$-$\sigma$ maximum BL voltage gives $536$~mV, increased to $646$~mV when accounting for the statistical SA offset. The upper bound is obtained from the worst-case BL leakage when both activated bitcells store $1$ while all others store $0$. Starting from $V_{\mathrm{DD}}$ gives $797$~mV, reduced to $781$~mV after SA offset uncertainty. The resulting $135$-mV margin at $V_\mathrm{DD} = 0.8$~V confirms reliable operation, with larger margins at higher supply voltages. A similar analysis on the BL side yields comparable results.

\paragraph{Basic bit-serial accumulation}\label{sec:bit_serial}

\begin{figure}
    \centering
    \includegraphics[width=\linewidth]{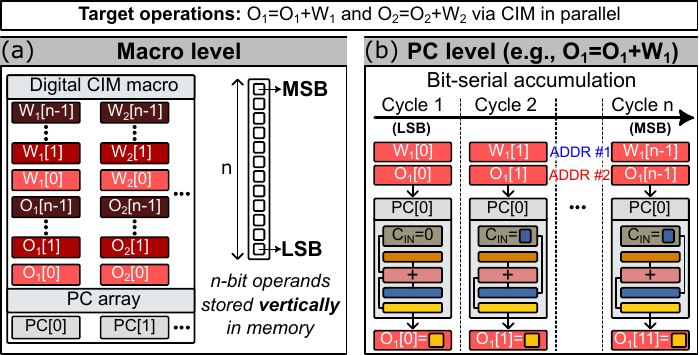}
    \caption{(a) Macro-level representation of operand locations in memory for a $n$-b weight-stationary bit-serial accumulation. (b) Temporal execution flow of the bit-serial accumulation $O_1=O_1+W_1$ and the involved PC modules.}
    \label{fig:bit_serial_accumulation}
    \vspace{-0.5cm}
\end{figure}

Before introducing the full operand-reshaping capabilities of FlexSpIM, we first describe a simple bit-serial accumulation situation in a WS configuration. Fig.~\ref{fig:bit_serial_accumulation}(a) shows two $n$-bit parallel accumulations, $O_{1}+W_{1}$ and $O_{2}+W_{2}$, executed by the first two PCs, respectively denoted $\mathrm{PC[0]}$ and $\mathrm{PC[1]}$. The operands are stored vertically along a bitline, going from the most-significant bit (MSB) at the top to the least-significant bit (LSB) at the bottom. During a bit-serial operation, one bit from each operand is processed per cycle, starting from the LSB and accessed from the full/half range of addresses $\mathrm{ADDR\#1/\#2}$ for the stationary/non-stationary operand (Fig.~\ref{fig:bit_serial_accumulation}(b)). The resulting sum bit overwrites the currently processed bit of $O_{1}$ and $O_{2}$. The generated carry is stored in a dedicated PC register and reused as the carry-in during the next cycle. OS behaves in a similar way, except that the operand positions in memory are swapped to have the weights at the bottom.

\paragraph{Operand reshaping for bit-parallel accumulation}\label{sec:operand_reshaping}

\begin{figure}
    \centering
    \includegraphics[width=\linewidth]{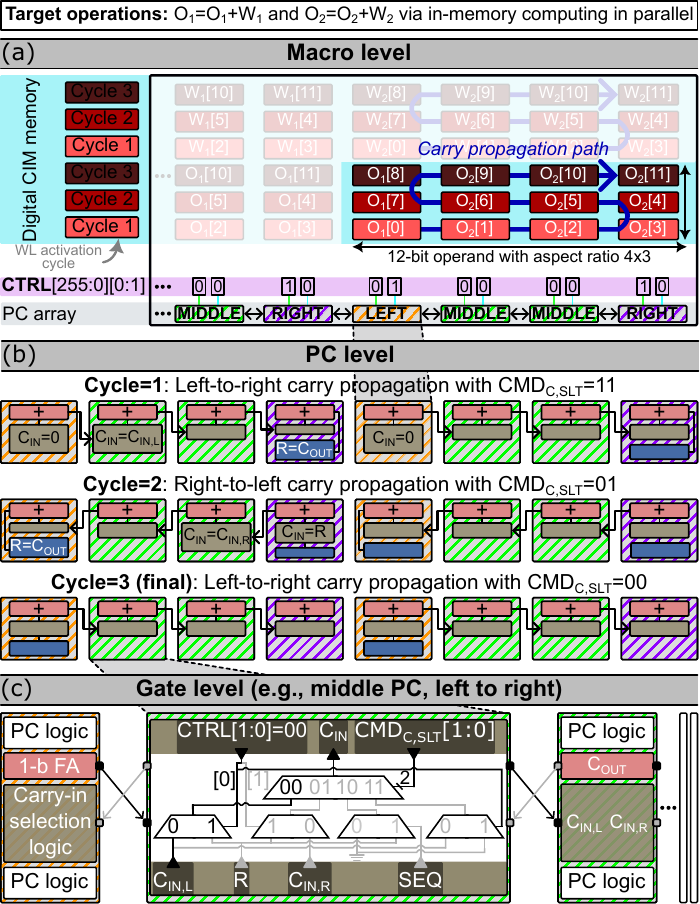}
    \caption{(a) Macro-level representation of operand locations in memory for a 12-bit weight-stationary bit-parallel accumulation. (b) Carry propagation during the accumulation process and the involved PC modules. (c) Detailed circuit implementation of the \textit{carry-in selection} PC module.}
    \label{fig:carry_propagation}
    \vspace{-0.5cm}
\end{figure}

To support flexible-precision bit-parallel accumulation without the storage penalty of interleaved-operand techniques \cite{IMPULSE, SPIDR}, we propose to directly reshape operands inside the memory (Fig.~\ref{fig:carry_propagation}). To illustrate the operating principle, we show a 12-bit addition of $O_{1}+W_{1}$ and $O_{2}+W_{2}$. The extension to different operand resolutions is discussed in Section~\ref{sec:overflow_underflow}. Each operand is stored as a $4\times3$ block spanning four columns and three rows. Bits are ordered left-to-right in even rows and right-to-left in odd rows.

Unlike bit-serial accumulation, all bits stored in a row of the SRAM array are processed in a single cycle, reducing the addition's latency from $12$ to $3$ cycles. To that end, we propagate full-adder (FA) carries between neighboring PCs, flowing left-to-right for even rows and right-to-left for odd ones (Fig.~\ref{fig:carry_propagation}(a)-(b)). The first PC in a block receives the resulting carry from the last cycle, initialized at $0$, intermediate PCs forward carries, and the last PC stores the new (intermediate) carry for the next cycle.

\begin{figure}
    \centering
    \includegraphics[width=\linewidth]{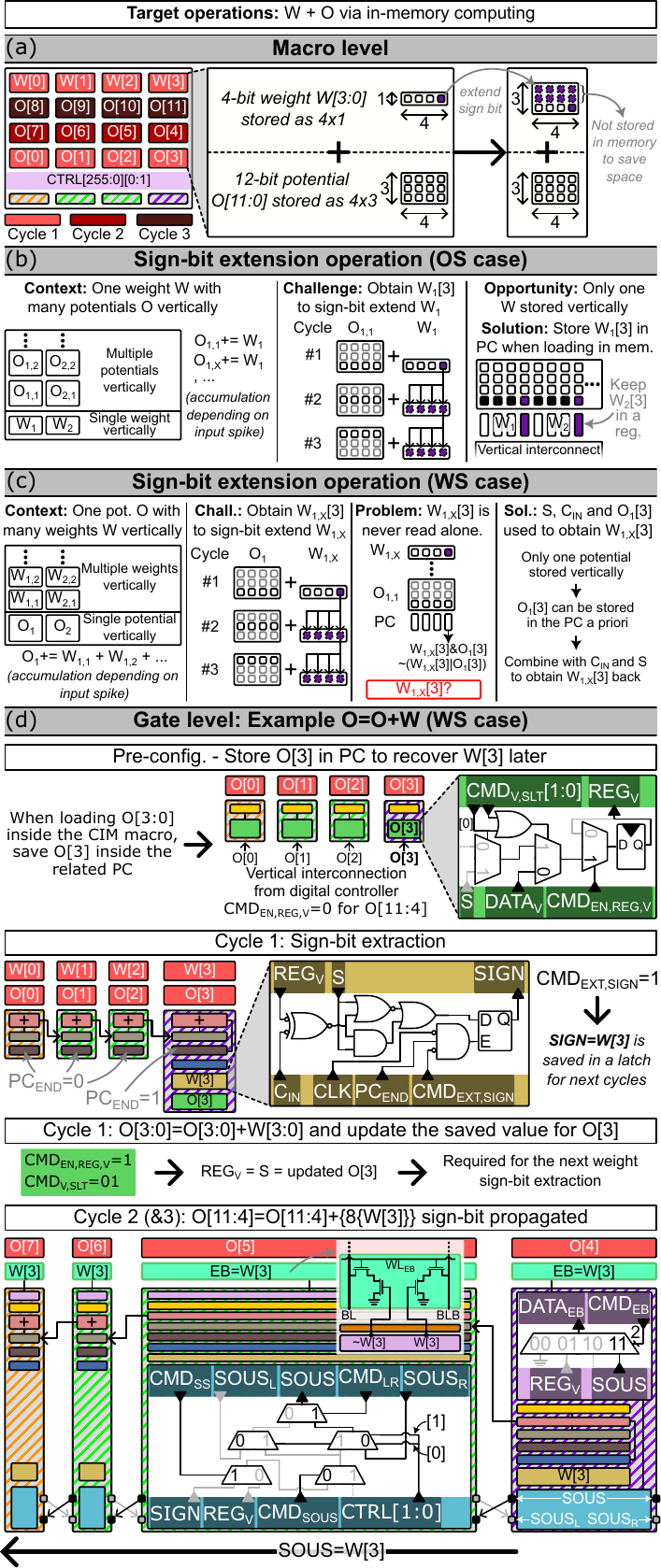}
    \caption{(a) Macro-level illustration of the sign-bit extension mechanism, with a 4-bit weight and 12-bit potential accumulation example. (b-c) Context, challenges, and proposed solution for sign-bit extension in the output-/weight-stationary (OS/WS) cases. (d) Cycle-level execution flow of sign-bit extension for WS, with detailed schematics of the relevant PC modules. $\mathrm{SOUS}$ propagates the weight \textbf{S}ign-bit, the \textbf{O}verflow and \textbf{U}nderflow indicators, and the \textbf{S}pike value to the current and neighboring PCs. $\mathrm{CMD}_{\mathrm{LR}}$ sets $\mathrm{SOUS}$ from either $\mathrm{SOUS_{L}}$ (left) or $\mathrm{SOUS_{R}}$ (right), while $\mathrm{CMD_{SS}}$ selects between $\mathrm{SIGN}$ and $\mathrm{REG_{V}}$ (spike storage). $\mathrm{PC_{END}}$ indicates if the current PC is the rightmost of a block and is computed from $\mathrm{CTRL[1:0]}$ locally.}
    \label{fig:sign_extension}
    \vspace{-1cm}
\end{figure}

To identify their first, middle, or last position in a block, or to disable them, PCs are configured by $\mathrm{CTRL[1:0]}$ signals, driven by the 2-bit control bitcells (Fig.~\ref{fig:carry_propagation}(c)). A global signal, $\mathrm{CMD_{C,SLT}[1:0]}$, selects the carry direction and initial carry-in, while $\mathrm{SEQ}$ enables a bit-serial accumulation. Contrary to prior works with fixed-width operands and static mapping patterns \cite{IMPULSE, SPIDR}, this configuration allows arbitrary block width and height, and therefore arbitrary operand resolution.

\subsubsection{Sign-bit extension}\label{sec:bit_sign_extension}

Until now, both operands were assumed to have the same resolution. However, in quantized neural networks, weights typically have a lower precision than accumulation, e.g., on membrane potentials for SNNs in Fig.~\ref{fig:sign_extension}(a). Assuming a 12-bit membrane potential and a 4-bit weight, the accumulation requires three CIM operations: the first performs a conventional multi-bit addition, while the remaining two perform weight sign extension.

Reading the sign bit before every accumulation would add an extra memory access and increase both energy consumption and latency by up to $50\%$ in the block-height-two membrane-potential configuration, a configuration primarily targeting speed. IMPULSE~\cite{IMPULSE} and SpiDR~\cite{SPIDR} instead avoid this overhead by fixing the corresponding membrane potential bit to 0 (Section~\ref{sec:soa_design}), reducing the membrane potential resolution by one bit and degrading the overall network accuracy. To avoid these limitations, the proposed sign-extension mechanism differs depending on whether OS or WS is used, as illustrated in Fig.~\ref{fig:sign_extension}(b) and Fig.~\ref{fig:sign_extension}(c-d), respectively. In both cases, the CIM operation is destructive because the weight sign bit (e.g., $W_1[3]$) cannot be distinguished from the membrane potential bit (i.e., $O_{1}[3]$), so we introduce ways to recover $W_1[3]$. For OS, only one weight is stored vertically and reused before being overwritten. Its sign bit is therefore stored directly in the corresponding PC register $\mathrm{REG_{V}}$ when the weight is loaded into memory (see Section~\ref{sec:operand_initialization}). For WS, multiple weights are stored simultaneously, making this approach impractical. Instead, FlexSpIM stores the membrane potential bit $O_{1,X}[3]$ in $\mathrm{REG_{V}}$ (Fig.~\ref{fig:sign_extension}(d), pre-configuration) and recovers the weight sign bit from the $\mathrm{S}$, $\mathrm{C_{IN}}$, and stored $O[3]$ signals when the last weight row is processed (Fig.~\ref{fig:sign_extension}(d), sign-bit extraction). The extracted sign bit is latched in $\mathrm{SIGN}$, while $\mathrm{REG_{V}}$ is updated with $S$ for subsequent accumulations. For both OS and WS, during the sign-extension cycles (Fig.~\ref{fig:sign_extension}(d), cycles 2--3), only potential rows are activated. The sign bit stored in $\mathrm{SIGN}$ or $\mathrm{REG_{V}}$ is propagated across the block PCs and emulated directly into the memory via the EBs (4T cells), thereby allowing to reuse the CIM operations.

\subsubsection{Overflow/Underflow protection mechanism}\label{sec:overflow_underflow}
\begin{figure}
\centering
\includegraphics[width=\linewidth]{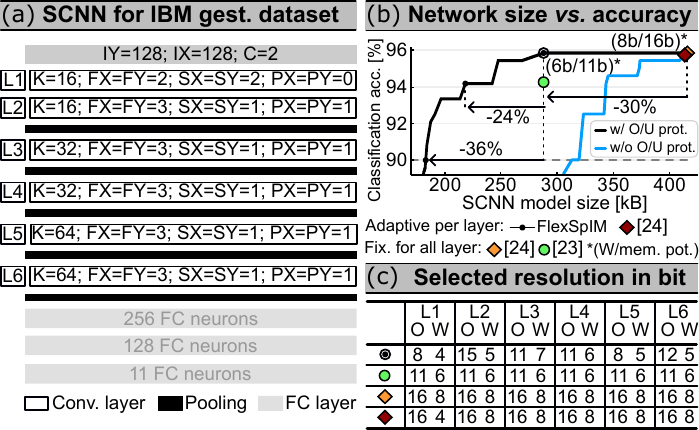}
\caption{(a) SCNN architecture for the IBM Gesture dataset, comprising six convolutional layers and three fully connected (FC) layers, the latter not being considered in this work. (b) Model size vs. accuracy achieved by FlexSpIM through configurable operand resolutions, compared with state-of-the-art architectures. The accuracy with and without overflow/underflow (O/U) are provided. (c) Per-layer operand resolutions of the points shown in (b).
}
\label{fig:network_size}
\vspace{-0.5cm}
\end{figure}

Reducing operand resolution decreases network size and external memory traffic. For the IBM Gesture SCNN in Fig.~\ref{fig:network_size}(a), the resolution flexibility of FlexSpIM (Fig.~\ref{fig:network_size}(c)) reduces convolutional-layer memory by up to $30\%$ without accuracy loss compared to analog CIM SNN accelerators~\cite{ISSCC24}. It also achieves higher accuracy than IMPULSE for a given network size, or the same accuracy with $24\%$ less memory.

However, aggressive quantization increases the risk of overflow/underflow (O/U), which can corrupt membrane potentials and degrade accuracy, reducing it from $95.8\%$ to less than $90\%$ at iso model size (Fig.~\ref{fig:network_size}(c)). While conventional digital accelerators often use saturation logic~\cite{clipping}, prior in-column digital CIM architectures~\cite{NEURALCACHE, SPIDR, IMPULSE} do not integrate O/U correction and instead avoid it by allocating sufficient numerical range. The challenge lies in performing fine-grained O/U detection and correction in CIM architectures due to strict pitch constraints. To address this challenge, FlexSpIM introduces a near-memory shared O/U detection unit and a fine-grained correction mechanism.

\begin{figure}
    \centering
    \includegraphics[width=\linewidth]{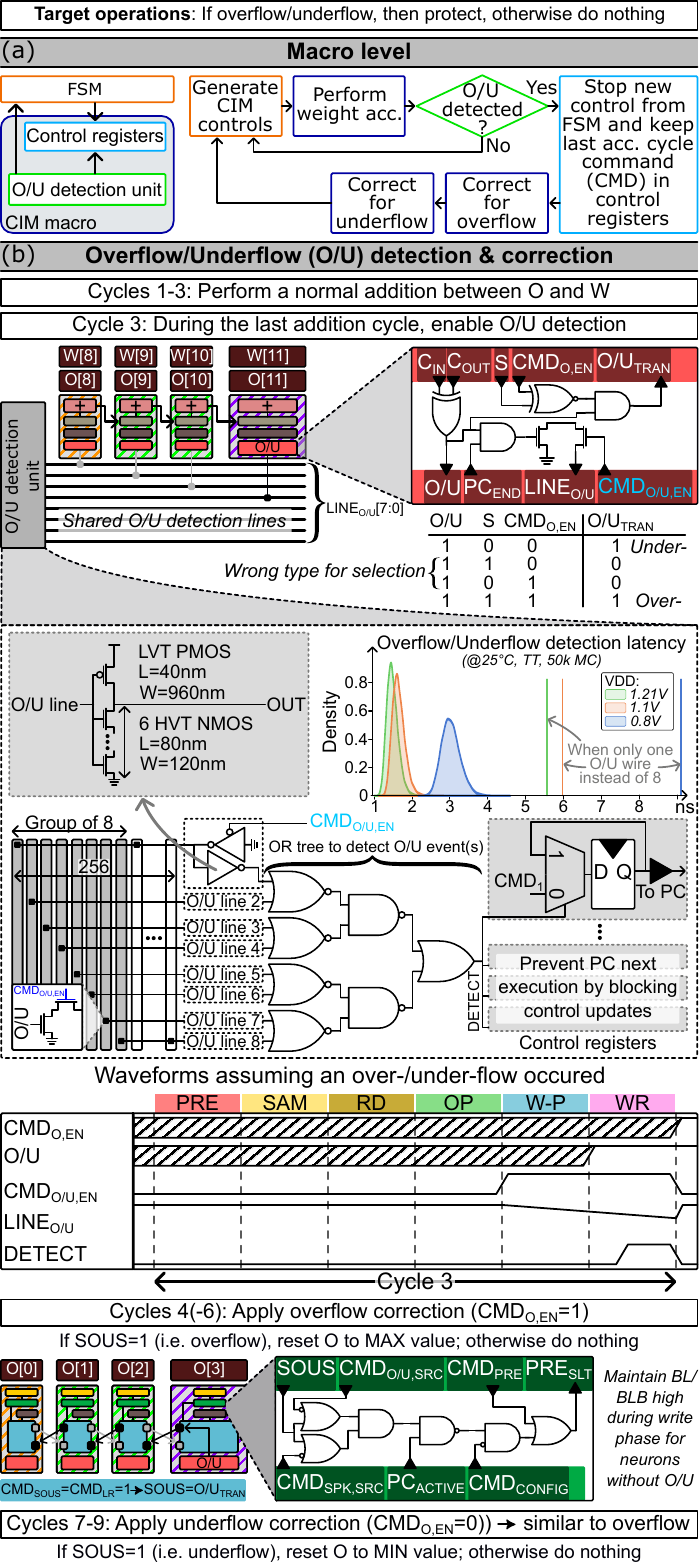}
    \caption{(a) Flow diagram of the proposed overflow/underflow detection and correction mechanism. (b) Cycle-accurate example of a 12-bit weight and membrane potential accumulation with an overflow or underflow, including the detailed circuitry of the O/U detection PC module and detection unit. $\mathrm{SOUS}$ stands for \textbf{S}ign-bit, the \textbf{O}verflow and \textbf{U}nderflow, and the \textbf{S}pike. $\mathrm{CMD_{O/U, SRC}}$ and $\mathrm{CMD_{SPK, SRC}}$ define whether the precharge select $\mathrm{PRE_{SLT}}$ is controlled by $\mathrm{SOUS}$ carrying the overflow/underflow signal or the spike value, respectively}
    \label{fig:overflow}
    \vspace{-0.5cm}
\end{figure}

The high-level principle is shown in Fig.~\ref{fig:overflow}(a). A normal accumulation is first completed. During its final cycle, an O/U detection unit determines whether any membrane potential overflowed or underflowed. If no event is detected, execution proceeds normally. Otherwise, the FSM stalls incoming commands while preserving the PC state and inserts two correction steps: one for overflow and one for underflow.

Fig.~\ref{fig:overflow}(b) details the O/U hardware execution for a 12-bit weight and membrane potential example. During the last accumulation cycle (e.g., cycle $3$), the MSB PC of each operand block detects O/U from its carry-in/-out, and sum bit. If detected, it discharges one of eight precharged $\mathrm{LINE_{O/U}}$ wires shared across the array, requiring up to $4$~ns at $V_\mathrm{DD}=0.8$V in the worst case (as opposed to $9.1$~ns, a $3\times$ longer average detection time $3\times$ longer if only one shared $\mathrm{LINE_{O/U}}$ were used). These wires are monitored by the O/U detection unit through high-threshold inverters. Once an O/U is detected, the correction phase is triggered. First, for overflow correction ($\mathrm{CMD_{O, EN}}=1$), the MSB PC propagates the O/U state ($\mathrm{O/U_{TRAN}}$) across the PCs of the block, using the same mux network as for sign-bit extension (Fig.~\ref{fig:sign_extension}(c), bottom). This produces $\mathrm{SOUS}$, a wire previously used in Section~\ref{sec:bit_sign_extension} to carry the weight sign-bit, now extended to also propagate the overflow and underflow signals, and later the spike signal (Section~\ref{sec:comparison}), to neighboring PCs. PCs with $\mathrm{SOUS}=1$ then overwrite the membrane potential for each of their block rows with a maximum/user-defined value provided through the vertical interconnect, while unaffected PCs remain in precharge mode through $\mathrm{PRE_{SLT}}$ thanks to the precharge-write phase avoiding half-select disturbance. Underflow ($\mathrm{CMD_{O, EN}}=0$) is handled identically. These corrections require the same number of cycles as the accumulation itself (i.e. three in the example), corresponding to a maximum CIM-cycle overhead of $200\%$. In practice, this overhead is hidden since system performance is typically limited by EMAs rather than by the CIM macro itself.

\subsubsection{Membrane-potential comparison and spike generation}\label{sec:comparison}

\begin{figure}
    \centering
    \includegraphics[width=\linewidth]{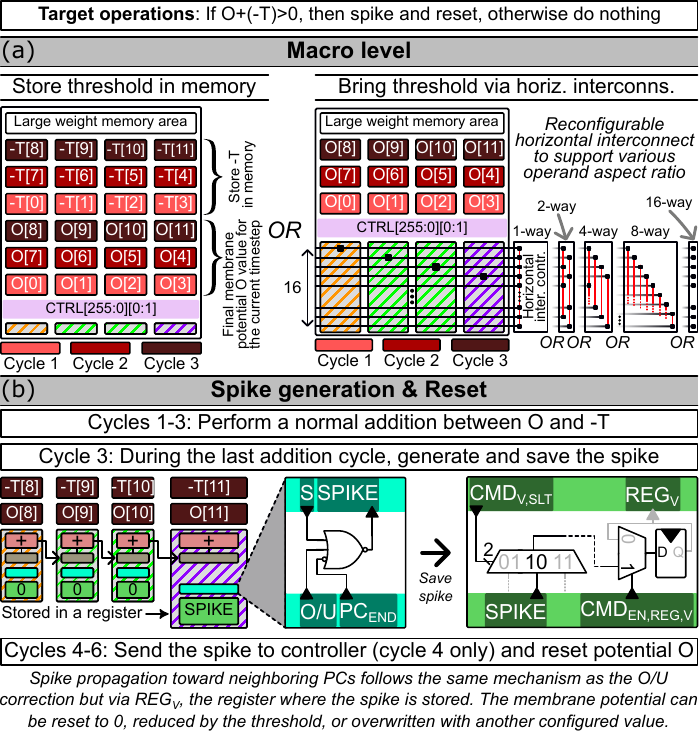}
    \caption{(a) Macro-level representation of two threshold provisioning methods: conventional in-memory storage (left) and the proposed shared horizontal interconnections (right). (b) Cycle-level representation of a threshold comparison with the corresponding PC module schematics.}
    \label{fig:spike_genaration_and_reset}
    \vspace{-0.3cm}
\end{figure}

When all spikes of a timestep have been processed, membrane potentials are compared against their threshold. While IMPULSE/SpiDR require thresholds to be stored inside the CIM macro, thereby causing memory overhead, FlexSpIM also supports threshold broadcasting through horizontal interconnects (Fig.~\ref{fig:spike_genaration_and_reset}(a)). The comparison then uses the EBs, originally introduced for sign extension, to emulate the threshold bits. This is beneficial when all neurons in a layer share the same threshold value~\cite{SAME_THRESHOLD}, saving $3$kb bit ($3\%$) of CIM storage for a fully bit-serial 16-bit potential/threshold configuration. The horizontal interconnect consists of $16$ wires, each connected to the PCs in an interleaved way with a stride of $16$. The interconnect controller supports 1-, 2-, 4-, 8-, and 16-way broadcast modes to match the operand shape. For example, fully bit-serial execution uses a 1-way broadcast, while a potential spanning two PCs uses a 2-way broadcast. Other operand shapes require either going back to threshold storage in memory or additional horizontal interconnect resources.

As shown in Fig.~\ref{fig:spike_genaration_and_reset}(b), spike generation follows the IMPULSE/SpiDR approach by adding the negative threshold to the membrane potential without overwriting it. An underflow indicates spike generation with the resulting $\mathrm{SPIKE}$ signal stored in $\mathrm{REG_V}$. When a neuron spikes, $\mathrm{SPIKE}$ is forwarded to the digital controller via the vertical interconnections and propagated across block PCs using the mux-network employed for sign-bit and O/U propagation. During reset, only PCs associated with $\mathrm{SPIKE}=1$ update their membrane potential.

\vspace{-0.3cm}
\subsection{Comparison to the State of the Art}\label{sec:comparison_table}

\begin{figure}
    \centering
    \includegraphics[width=\linewidth]{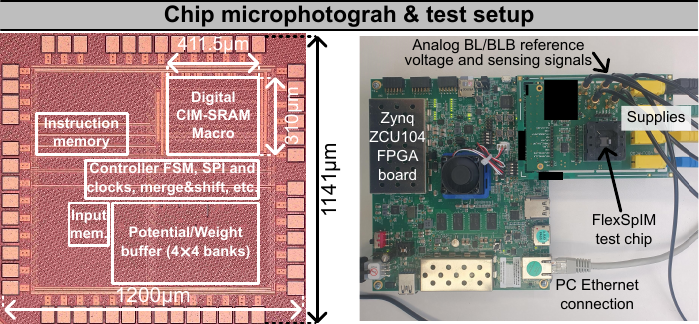}
    \caption{Chip microphotograph and test setup.}
    \label{fig:microphotograph_and_pcb}
    \vspace{-0.5cm}
\end{figure}

\begin{table*}[!htbp]
\caption{Comparison to the state of the art of SNN accelerators.}
\vspace{-0.5cm}
\begin{center}
\begin{footnotesize}
\resizebox{\textwidth}{!}{
\begin{tabular}{|C{0.21\linewidth}!{\color{blue}\vrule width 0.4mm}C{0.095\linewidth}!{\color{blue}\vrule width 0.4mm}C{0.095\linewidth}|C{0.135\linewidth}|C{0.095\linewidth}|C{0.095\linewidth}|C{0.11\linewidth}|C{0.105\linewidth}|}
\arrayrulecolor{black}
\cline{1-1}\cline{3-8}
\arrayrulecolor{blue}\cline{2-2}
\arrayrulecolor{black}

\textbf{} & \textbf{This work} & SSC-L'21~\cite{IMPULSE} & arXiv'24~\cite{SPIDR} & ISSCC'24~\cite{ISSCC24} & JSSC'23~\cite{NEUROCIM} & A-SSCC'22~\cite{SPIKECIM}& ISSCC'22~\cite{ReckOn}\\
\hline
Technology & 40nm & 65nm & 65nm & 22nm & 28nm & 65nm & 28nm\\
Implementation & \textbf{Digital (CIM)} & \textbf{Digital (CIM)} & \textbf{Digital (CIM)} & Analog CIM & Analog CIM & Analog CIM & Digital\\
Core area (mm$^2$) & 1.37 & 0.089 $^{\textbf{a}}$ & 3.12 & 2.28 & 2.9 & 0.25 $^{\textbf{a}}$ & 0.45\\
Macro memory capacity (kB) & 16 & \textbf{\textcolor{red}{1.37}} & 9.7 & 4 & \textbf{\textcolor{blue}{20}} & 4 & N/A\\
Bitcell type & 6T & 10T & 10T & 6T & 8T & 2$\times$6T+6T & N/A\\
\hline
Spiking network type & CNN & Fixed CNN & Fixed CNN & Residual CNN & Residual CNN & CNN & RNN\\
Representative dataset & DVS gesture $^{\textbf{b}}$ & MNIST/IMDB & DVS gesture/DSEC & DVS gesture $^{\textbf{b}}$ & CIFAR-10 & MNIST/CIFAR10 & DVS gesture $^{\textbf{b}}$\\
Accuracy on DVS gesture & \textcolor{blue}{$95.8\%$} & N/A & $88.44\%$ & $94\%$ & N/A & N/A & \textcolor{red}{$87.3\%$}\\
\hline
Multi-aspect ratio support & \textbf{\textcolor{blue}{\checkmark}} & \textcolor{red}{$\times$} & \textcolor{red}{$\times$} & \textcolor{red}{$\times$} & \textcolor{red}{$\times$} & \textcolor{red}{$\times$} & \textcolor{red}{$\times$}\\
OS/WS support & \textbf{\textcolor{blue}{\checkmark}} & \textcolor{red}{$\times$} & \textcolor{red}{$\times$} & \textcolor{red}{$\times$} & \textcolor{red}{$\times$} & \textcolor{red}{$\times$} & \textcolor{red}{$\times$}\\
Mem. pot. resolution & \textbf{\textcolor{blue}{Any}} & 11b & 7b/11b/15b & 16b & \textbf{\textcolor{red}{8b}} & Analog & 16b\\
Weight resolution & \textbf{\textcolor{blue}{Any}} & 6b & 4b/6b/8b & 4/8b & 1/4/8b & \textbf{\textcolor{red}{1.5b}} & 8b\\
Weight/Mem. pot. location & \textbf{\textcolor{blue}{Not fixed}} & \textbf{\textcolor{red}{Fixed}} & \textbf{\textcolor{red}{Fixed}} & \textbf{\textcolor{red}{Fixed}} & \textbf{\textcolor{red}{Fixed}} & \textbf{\textcolor{red}{Fixed}} & \textbf{\textcolor{red}{Fixed}}\\
\hline
Supply range (V) & 0.9 -- 1.1 & 0.7 -- 1.2 & 0.9 -- 1 (1.2) & 0.55 -- 0.9 & 1.1 & N/A & 0.5 -- 0.8\\
Frequency (MHz) & 75.5 -- 157 & 66.7 -- \textbf{\textcolor{blue}{500}} & 50 -- 150 & 51 -- 280 & 200 & N/A & \textbf{\textcolor{red}{13}} -- 115\\
Peak throughput (GSOPS)$^{\textbf{*}}$ & 1.2 -- 2.5 $^{\textbf{c}}$ & 0.07 -- 0.5 $^{\textbf{d}}$ & 0.9 -- 2.7 $^{\textbf{d}}$ & N/A & N/A & \textbf{\textcolor{blue}{163.8}} $^{\textbf{f}}$ & \textbf{\textcolor{red}{0.013}} -- 0.115 $^{\textbf{c}}$\\
1b-norm. throughput (GSOPS)$^{\ddag}$ & 154 -- \textbf{\textcolor{blue}{320}} & 4.62 -- 33 & 59.4 -- 178.2 & N/A & N/A & N/A & \textbf{\textcolor{red}{1.67}} -- 14.7\\
Power (mW) & 6.8 -- 17.9 $^{\textbf{a}}$ & \textbf{\textcolor{blue}{0.1}} -- 0.9 $^{\textbf{a}}$ & 4.9 -- \textbf{\textcolor{red}{18}} & 0.524 -- 6.4 & 15.84 $^{\textbf{a}}$ & 0.56 $^{\textbf{a}}$ & 0.077 -- N/A\\
Efficiency (pJ/SOP)$^{\textbf{*}}$ & 5.7 -- 7.2 $^{\textbf{c}}$ & 1.09 -- 1.74 $^{\textbf{d}}$ & 1.11 -- 1.36 $^{\textbf{d}}$ & 3.78 -- 10.01 $^{\textbf{c}}$ & \textbf{\textcolor{blue}{0.0016}} $^{\textbf{e}}$ & 3.45$\times$10$^{-3~ \textbf{f}}$ & 5.3 -- \textbf{\textcolor{red}{12.8}} $^{\textbf{c}}$\\
1b-norm. eff. (fJ/SOP)$^{\dag}$ & 44.5 -- 56.3 & 16.5 -- 26.4 & 16.8 -- 20.6 & 29.5 -- 78.2 & \textbf{\textcolor{blue}{0.025}} & N/A & 41.4 -- \textbf{\textcolor{red}{100}}\\
\arrayrulecolor{black}
\cline{1-1}\cline{3-8}
\arrayrulecolor{blue}\cline{2-2}
\arrayrulecolor{black}
\multicolumn{7}{l}{
\rule{0pt}{3mm}%
$^{\mathrm{\textbf{a}}}$ CIM macro only \quad
$^{\mathrm{\textbf{b}}}$ 10 classes \quad
$^{\mathrm{\textbf{c}}}$ 8-bit weight and 16-bit mem. pot. \quad
$^{\mathrm{\textbf{d}}}$ 6-bit weight and 11-bit mem. pot. \quad
$^\mathrm{\textbf{e}}$ 8-bit weight and 8-bit mem. pot.}\\
\multicolumn{7}{l}{$^{\mathrm{\textbf{f}}}$1.5-bit weight and N/A for mem. pot. \quad
$^{\mathrm{\ddag}}$GSOPS $\times$ weight-bit $\times$ pot.-bit \quad
$^{\mathrm{\dag}}$fJ/SOP/(weight-bit $\times$ pot.-bit) \quad
$^{\mathrm{*}}$1 SOP = 1 addition + mem. pot. update}\\
\end{tabular}
}
\end{footnotesize}
\label{tab:comparison_table}
\end{center}
\vspace{-0.4cm}
\end{table*}

Considering the CIM macro fabricated in TSMC 40~nm, with its microphotograph and test setup shown in Fig.~\ref{fig:microphotograph_and_pcb}, Table~\ref{tab:comparison_table} compares FlexSpIM against existing architectures. FlexSpIM provides the highest flexibility through bitwise resolution reconfiguration and support for two dataflows, while achieving $5\times$ higher throughput with a 1-bit-normalized energy efficiency no more than $\sim2\times$ lower than the best digital CIM architectures. \cite{NEUROCIM} achieves three orders of magnitude higher efficiency thanks to the use of analog CIM but only supports an 8-bit membrane potential with low flexibility. Although the higher flexibility of FlexSpIM comes at the cost of lower energy efficiency, we show in Section~\ref{sec:results} that this penalty becomes negligible at the system level, where flexibility is a much more important driver to system-level efficiency than macro-level energy metrics.

%% file: text/system_level.tex
\vspace{-0.1cm}
\section{System-level Design Space Exploration and Software Stack} \label{sec:system}
The strength of FlexSpIM does not lie in maximizing standalone CIM macro-level efficiency, as increased flexibility inevitably incurs energy and latency overheads compared to rigid architectures. Instead, its key advantage lies in exploiting this flexibility to reduce data movement at the system level. This tradeoff is particularly important, as Houshmand \textit{et al.}~\cite{ACIM_ZigZag} showed that a CIM macro achieving 5.9~fJ/op suffered a $10\times$ system-level energy efficiency loss due to data transfers between the CIM macro, on-chip buffers, and off-chip memory. When the complete weight set could not be stored on chip, an additional $2\times$ energy penalty was reported. In this scenario and for SCNNs executed with a layer-first schedule (Section~\ref{sec:snn_execution}), sparse activations further increase the overhead of data movement as weight reuse is reduced.

\vspace{-0.3cm}
\subsection{Hybrid-Stationary Dataflow}
\begin{figure*}[t]
    \centering
    \includegraphics[width=\linewidth]{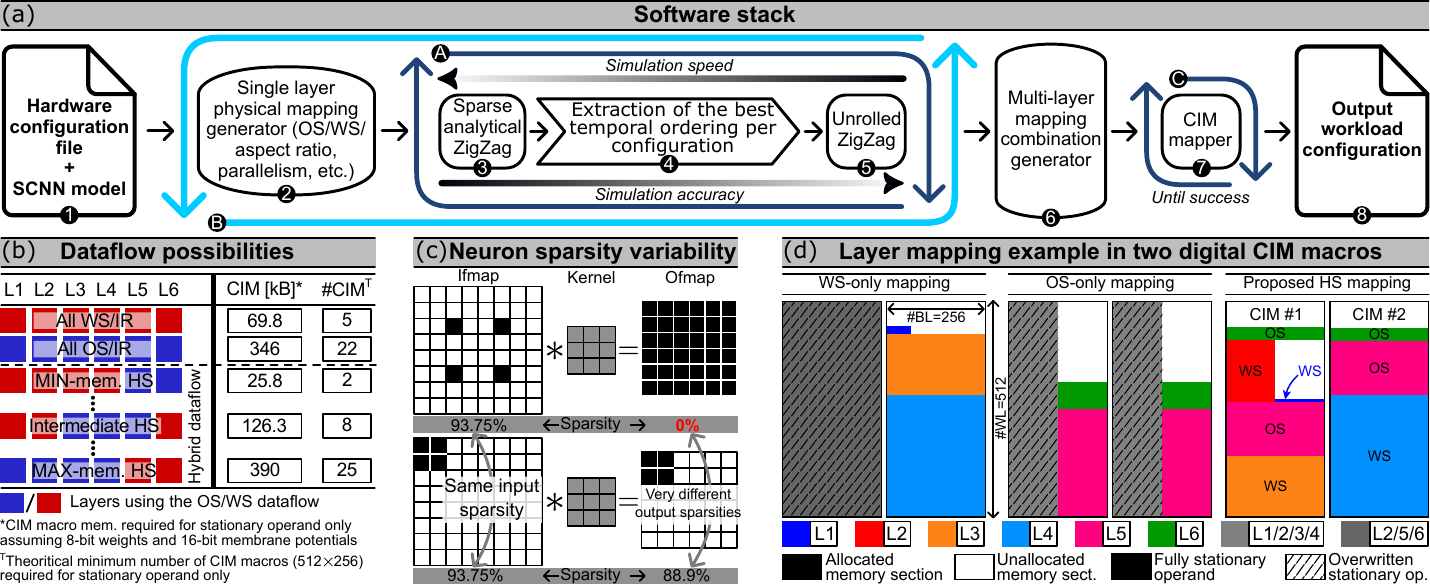}
    \caption{(a) Software stack selecting the OS/WS mapping, parallelization, and operand block dimensions for each workload layer to optimize energy and latency under a fixed (multi-)macro FlexSpIM configuration. (b) Supported basic and hybrid stationarity configurations in FlexSpIM, with the required CIM memory size and macro count assuming ideal utilization. (c) Impact of sparse input locations on the number of affected output neurons. (d) Mapping generated by the software stack in (a) of the six SCNN layers onto two CIM macros for hybrid stationarity, OS-only, and WS-only execution.}
    \label{fig:software_stack}
    \vspace{-0.4cm}
\end{figure*}

Unlike IMPULSE, which supports only WS execution, FlexSpIM enables hybrid stationarity (HS) on a per-layer basis, where each layer independently adopts either WS or OS depending on the available CIM memory. For the SCNN in Fig.~\ref{fig:network_size}(a), storing all weights requires $69.8$~kB, corresponding to a minimum of five $256\times512$ CIM macros assuming ideal utilization (Fig.~\ref{fig:software_stack}(b)). With fewer macros, a WS-only architecture repeatedly reloads weights, substantially increasing memory traffic. In contrast, the proposed HS requires only two macros to keep one operand of each layer permanently inside the CIM array, referred to as the \textit{min-memory HS} configuration. With enough macros, the \textit{max-memory HS} keeps stationary the operand with the largest memory footprint for each layer, further reducing data movement. Intermediate configurations are also supported, illustrating a broad search space~\cite{ZigZag}, shaped by application requirements.

To efficiently explore this large design search space, we introduce the software stack shown in Fig.~\ref{fig:software_stack}(a). The software takes as input \circled[0.05em]{1} the target hardware configuration, including memory hierarchy, capacity, bandwidth, access energy, and number of CIM macros. For each layer (loop \circled[0.05em]{B}), it evaluates multiple execution configurations (loop \circled[0.05em]{A}, Section~\ref{sec:single_layer}) and generates candidate hardware mappings. These mappings are then jointly optimized across CIM macros (loop \circled[0.05em]{C}), Section~\ref{sec:multi_layer}, producing the final workload mapping with its estimated energy and latency.

\vspace{-0.3cm}
\subsection{Single-Layer Performance Exploration}\label{sec:single_layer}
To explore the layer-level design space efficiently, we split it into two components: \textit{physical mappings} and \textit{intra-layer scheduling}. \textit{Physical mappings} define how a layer is spatially mapped onto CIM macros, including dataflow choice (OS/WS), parallelization level, and operand shaping (bit-serial, bit-parallel, or mixed). These choices directly shape the mapping geometry: higher parallelization and bit-parallel execution enable wider horizontal mappings but increase on-chip and off-chip memory bandwidth pressure, whereas lower parallelization with bit-serial execution enables more vertical mappings at the potential cost of higher latency. Candidate mappings that exceed available CIM macro dimensions are also considered as fallback configurations when full stationarity cannot be maintained. All these configurations are generated by the physical mapping generator \circled[0.05em]{2}, which outputs configuration files to be used by the system performance simulator \circled[0.05em]{3}. The objective of this \textit{physical mapping} space exploration is not to maximize the performance of individual layers, but to enable efficient co-location of all layers within the same CIM macro set and optimize workload-level performance. Therefore, generating diverse mapping shapes is essential, as a mapping optimal for one layer may hinder efficient integration of the remaining workload's layers, as discussed in Section~\ref{sec:multi_layer}.

The second component is \textit{intra-layer scheduling}, which defines the temporal execution order within a fixed physical mapping. It does not modify the spatial placement of a layer inside the CIM macro, but determines execution order, e.g., processing all input spikes of a neuron before moving to the next neuron, or processing all neurons affected by a spike before proceeding to the next spike, among other strategies~\cite{ZigZag}. To evaluate these two design space components, we rely on two simulators: an \textit{analytical} one and a \textit{cycle accurate} one.

\subsubsection{Analytical ZigZag}\label{sec:analytical_zigzag}
The first simulator \circled[0.05em]{3} builds on the design-space exploration tool ZigZag~\cite{ZigZag} and targets the \textit{intra-layer scheduling} space. For each \textit{physical mapping} (loop \circled[0.05em]{A}), it evaluates candidate schedules using a fast analytical energy model for data transfers and CACTI~\cite{CACTI} for memory accesses. Fast models are essential to explore the large design space, which can reach billions of possibilities for our SCNN across scheduling, depending on the representation granularity~\cite{ZigZag}.

To support SCNN execution, we extend the analytical model proposed in~\cite{ZigZag} to include the timestep dimension, membrane potential persistence across timesteps in a layer-first configuration, and input sparsity, under a homogeneous spike distribution assumption. While this assumption preserves computational efficiency for fast evaluation, it reduces estimate accuracy on energy/latency compared to the deterministic dense workloads targeted by the original ZigZag framework. In practice, spike distributions can be highly non-uniform: as shown in Fig.~\ref{fig:software_stack}(c), for the same input sparsity (e.g., $93.75\%$), spatially localized spikes activate only $11\%$ of output neurons (i.e., $89\%$ output sparsity), whereas uniformly distributed spikes activate all output neurons ($0\%$ output sparsity). The homogeneous assumption represents a worst-case scenario: although the total number of synaptic operations remains unchanged, it upper-bounds the number of membrane potentials that must be fetched from memory into the CIM macro.

\subsubsection{Unrolled (cycle-accurate) ZigZag}\label{sec:unrolled_zigzag}
To improve modeling accuracy, we introduce a second simulator \circled[0.05em]{5}, referred to as \textit{unrolled ZigZag}. Unlike \textit{analytical ZigZag}, it performs cycle-accurate execution by explicitly unrolling the full workload on the real input data, yielding more accurate latency and energy estimates at the cost of significantly longer execution time. This makes exhaustive evaluation of all scheduling and mapping combinations impractical. Since \textit{intra-layer scheduling} does not affect the physical placement of a layer inside CIM macros, only the best schedule per \textit{physical mapping}, obtained from \circled[0.05em]{4}, is required in multi-layer mapping (Section~\ref{sec:multi_layer}). \textit{Analytical ZigZag} thus identifies the optimal intra-layer schedule per \textit{physical mapping}, while \textit{unrolled ZigZag} explores only the \textit{physical mapping} space. This reduces the design space from billions of configurations to a few hundred evaluations. 

\vspace{-0.3cm}
\subsection{Multi-Layer Mapping Optimizer}\label{sec:multi_layer}

After selecting the best schedule for each physical mapping, the final step is to determine the combination of layer mappings that optimizes workload-level energy and latency. The mapping configuration generator \circled[0.05em]{6} explores combinations by selecting one physical mapping per layer, ordered by their individual performance, and provides them to the CIM mapper \circled[0.05em]{7}, which verifies their feasibility on the target macro array. It stops as soon as one combination is found \circled[0.05em]{8}. This process is formulated as a rectangle packing problem, where layer mappings are rectangles to be placed inside the available CIM macro array without overlap. Since this problem is NP-complete~\cite{NP_complete}, we employ the MaxRects Bottom-Left heuristic, which greedily places rectangles while maintaining maximal free regions~\cite{MaxRec}. Fig.~\ref{fig:software_stack}(d) compares the resulting mappings for the six-layer SCNN from Fig.~\ref{fig:network_size}(a) using two CIM macros. While WS-only and OS-only approaches require some layers to break operand stationarity, hybrid stationarity maintains fully stationary execution for all layers.

%% file: text/results.tex
\section{Experimental Results and System-Level Analysis}\label{sec:results}

\begin{figure}
    \centering
    \includegraphics[width=\linewidth]{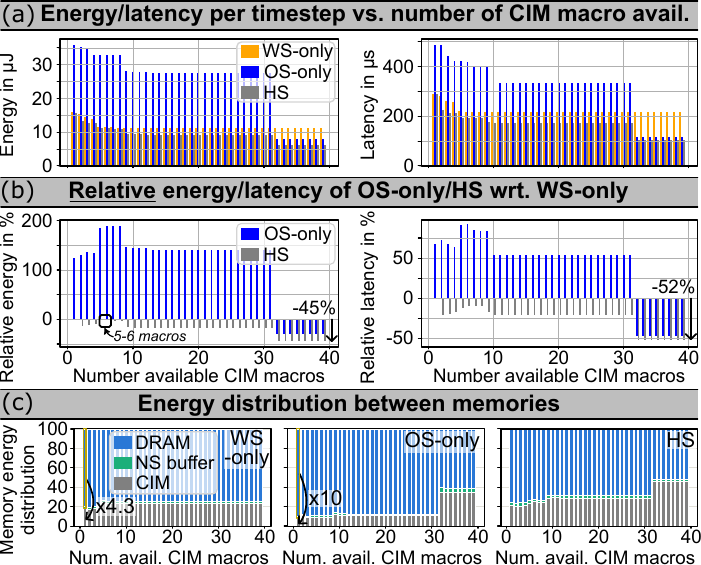}
    \caption{(a) Energy and latency obtained on average per executed timestep, for the SCNN in Fig.~\ref{fig:network_size}(a) on the IBM DVS gesture dataset, across systems composed of 1 to 39 CIM macros, for hybrid stationarity, OS-only, and WS-only executions. (b) Relative energy and latency of hybrid stationarity and OS-only execution with respect to WS-only execution (e.g., IMPULSE~\cite{IMPULSE} and SpiDR~\cite{SPIDR}) for the same configuration as (a). (c) Energy breakdown across the memories of the system, for the three dataflows. The input buffer and input scratches represent less than $0.1\%$ and are not shown.}
    \label{fig:energy_latency_results}
    \vspace{-0.5cm}
\end{figure}

\begin{figure}
    \centering
    \includegraphics[width=\linewidth]{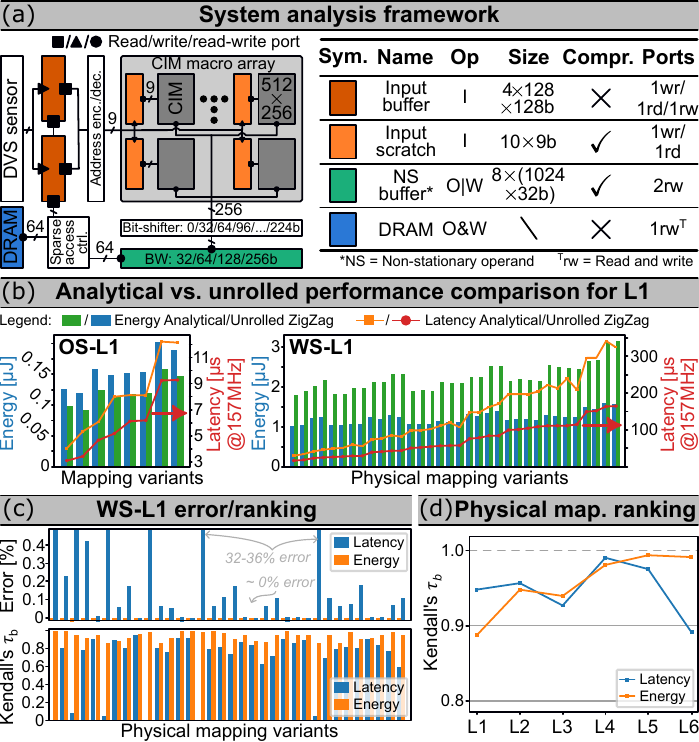}
    \caption{(a) System analysis framework, based on the measured CIM macro. (b) Energy and latency comparison between \textit{unrolled} and \textit{analytical ZigZag} for Layer~1 of the SCNN in Fig.~\ref{fig:network_size}(a), using the \textit{analytical-ZigZag}-selected schedule for each physical mapping under OS (left) and WS (right) execution. (c) Performance gap between the \textit{analytical-ZigZag}-selected schedule and the best schedule obtained from \textit{unrolled ZigZag} both using \textit{unrolled} results (bottom), and Kendall's $\tau_{b}$ correlation between \textit{analytical} and \textit{unrolled ZigZag} rankings for $100$ representative schedules of each WS physical mapping of Layer~1 (top). (d) Kendall's $\tau_{b}$ correlation between \textit{analytical} and \textit{unrolled ZigZag} physical-mapping rankings across all layers of the vanilla SCNN.}
    \label{fig:analytical_unrolled_comparison}
    \vspace{-0.5cm}
\end{figure}

\subsection{Experimental Setup}
Using the software stack described in Section~\ref{sec:system}, we evaluate the hardware platform shown in Fig.~\ref{fig:analytical_unrolled_comparison}(a) with 16-bit membrane potentials and 8-bit weights. The system consists of an external DRAM large enough to store the complete network parameters, on-chip input and non-stationary (NS) operand buffers, and a configurable CIM macro array based on the macro prototyped in 40nm CMOS presented in Section~\ref{sec:macro}. The input buffer stores 1-bit uncompressed DVS-generated spikes and provides dedicated ports for spike writing, CIM access, and generated spike storage. Depending on the selected dataflow, the NS buffer stores either membrane potentials (WS) or weights (OS). For WS execution, the input buffer includes a dedicated read port to process input spikes and identify the sparse membrane potentials that must be fetched from DRAM and stored in the NS buffer. The NS buffer supports adaptive bandwidths (32/64/128/256-bit) depending on the layer physical mapping width and connects to the CIM array through a 256-bit single-cycle transfer bus with 32-bit granularity shifting to support arbitrary mapping positions. Data stored in the NS buffer are pre-arranged according to their final CIM macro allocation. Each CIM macro is preceded by an input FIFO storing operand addresses generated by the spike-to-address encoder/decoder.

\vspace{-0.3cm}
\subsection{Results}
Fig.~\ref{fig:energy_latency_results}(a) presents the system-level energy and latency of OS-only, WS-only, and HS approaches for a single $10$~ms timestep across all layers of the SCNN in Fig.~\ref{fig:network_size}(a), on the DVS128 IBM gesture dataset~\cite{DVS128}, as a function of the number of CIM macros. Fig.~\ref{fig:energy_latency_results}(b) shows the relative improvements of OS-only and HS over WS-only (e.g., IMPULSE~\cite{IMPULSE}). HS achieves up to $52$\% latency and $45$\% energy reduction with $32$ macros, reaching $105\mu$s and $6.4\mu$J per timestep at $157$~MHz, versus $291\mu$s and $15.8\mu$J with a single macro. The lower HS gain around 5 macros matches the number of macros WS requires to store all weights on-macro (Fig.~\ref{fig:software_stack}(b)), making HS and WS-only performance close. The sharp OS improvement around 31 macros corresponds to L1 and L2 membrane potentials becoming fully stationary, i.e., stored entirely on-macro; beyond this point, no further gain occurs, since each layer's largest memory-consuming operand remains on-macro. While improvements depend on workload and model architecture, HS consistently offers the best trade-off across macro budgets via layer-level stationarity adaptation. Fig.~\ref{fig:energy_latency_results}(c) shows the energy distribution across the system's memories. For a single macro in WS-only, off-macro transfers account for $4.3\times$ more energy than the CIM macro itself, still $4\times$ more at 39 macros, confirming Houshmand \textit{et al.}~\cite{ACIM_ZigZag} observations. This shows that system-level optimizations, such as the proposed hybrid dataflow, matter more than further CIM macro optimization.

\vspace{-0.3cm}
\subsection{Discussion}
Since exhaustive cycle-accurate simulation of all candidate schedules is computationally prohibitive, Fig.~\ref{fig:analytical_unrolled_comparison}(b) compares the energy and latency estimated by \textit{analytical} and \textit{unrolled ZigZag} for the schedules identified by \textit{analytical ZigZag} across all physical mappings of L1 under OS and WS execution. As expected, larger absolute differences appear for WS, since OS keeps membrane potentials inside the CIM macros and is therefore less sensitive to spike distribution, highlighting the need for \textit{unrolled ZigZag} for accurate cost estimation. This approach is valid only if \textit{analytical ZigZag} preserves the optimal schedule under cycle-accurate evaluation.

Since a single physical mapping can contain billions of candidate schedules~\cite{ZigZag}, we validate this assumption on L1 by sampling 100 representative schedules per physical mapping and evaluating them with both simulators. Fig.~\ref{fig:analytical_unrolled_comparison}(c) compares, for each physical mapping, the performance obtained from \textit{unrolled ZigZag} using the \textit{analytical-ZigZag}-selected schedule, with the best performance obtained among the 100 sampled schedules evaluated by \textit{unrolled ZigZag}. Most physical mappings achieve near-optimal performance, with only six showing larger latency deviations, up to $36$\%. This agreement is facilitated by the NS buffer, which stores compressed membrane potentials during WS execution and reduces the impact of the homogeneous spike distribution assumption by limiting incorrect costly DRAM accesses.

A stronger validation criterion is whether \textit{analytical ZigZag} preserves the relative ranking of candidate schedules, beyond only identifying the best one. This demonstrates that the analytical model captures the dominant hardware bottlenecks and provides confidence that the selected schedules remain relevant outside the sampled cases. This ranking consistency is quantified using Kendall's $\tau_{b}$\cite{Kendall}, given by
\begin{equation}
    \tau_{b}=\frac{C-D}{\sqrt{(C+D+T_{x})(C+D+T_{y})}},
\end{equation}
where $C$ and $D$ denote the number of concordant and discordant pairs, respectively, $T_{x}$ and $T_{y}$ denote the numbers of tied pairs in the first and second rankings, respectively, and $\tau_{b}=1$ indicates identical rankings between both simulators. Fig.~\ref{fig:analytical_unrolled_comparison}(c) shows consistently high $\tau_{b}$ values across physical mappings, except for the six previously identified latency-sensitive cases. Removing pairs with negligible performance differences (below 1\%) improves the correlation, confirming that \textit{analytical ZigZag} preserves candidate schedule ordering despite the homogeneous spike distribution assumption.

Repeating the 100-schedule comparison for every layer is computationally prohibitive. Instead, we compute Kendall's $\tau_b$ on the physical-mapping ranking using the already obtained \textit{unrolled ZigZag} evaluations of the \textit{analytical-ZigZag}-selected schedules, Fig.~\ref{fig:analytical_unrolled_comparison}(d). The consistently high correlation (i.e., higher than $88\%$) across the workload confirms that \textit{analytical ZigZag} captures the dominant hardware trends beyond a single layer, validating our two-stage exploration methodology.

%% file: text/conclusion.tex
\section{Conclusion}\label{sec:conclusion}
This work introduces FlexSpIM, a flexible digital CIM architecture for mapping sparse spiking neural networks with diverse layer shapes onto memory-constrained accelerators. Unlike weight-stationary-only approaches, FlexSpIM enables layer-level hybrid weight or output stationarity to reduce operand movement. Operand reshaping inside the CIM macro supports arbitrary resolutions, reducing model size by $30\%$ compared to state-of-the-art CIM approaches while maintaining $95.8\%$ accuracy on the IBM DVS gesture dataset, while built-in overflow/underflow detection and correction enables further compression with limited accuracy loss. Fabricated in TSMC 40~nm CMOS technology, FlexSpIM achieves competitive performance with recent digital SNN accelerators despite its increased flexibility. Combined with a dedicated software stack for dataflow, scheduling, and physical mapping optimization, it improves system-level energy efficiency and latency by up to $45\%$ and $52\%$, respectively, over rigid stationarity approaches. These results show future CIM accelerators should optimize not only the CIM macro, but also provide sufficient flexibility to adapt to various workload requirements.

%% file: text/acknowledgments.tex
\vspace{-0.3cm}
\section*{Acknowledgments}
The authors thank Dr. Martin Lefebvre, Dr. Christoph Posch, Dr. Damien Joubert, and Prof. Kofi Makinwa for their valuable feedback and Lukasz Pakula for his help with die photographs and PCB design.

%% file: text/biography.tex
\vspace{-1.5cm}

\begin{IEEEbiography}[{\includegraphics[width=1in,height=1.25in,clip,keepaspectratio]{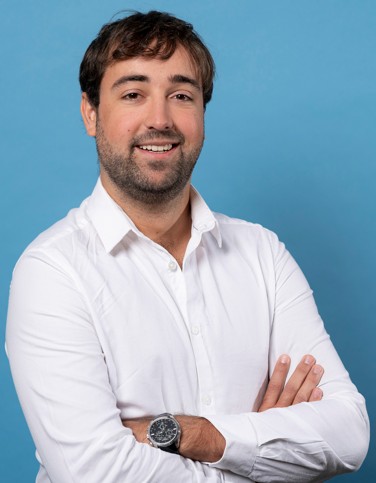}}]{Nicolas Chauvaux}
(Graduate Student Member, IEEE) received the B.Sc. degree (Magna cum laude) and M.Sc. degree (summa cum laude) in electrical engineering from the Catholic University of Louvain-la-Neuve (UCL), Louvain-la-Neuve, Belgium, in 2019 and 2021, respectively. He is currently pursuing a Ph.D. degree at the Delft University of Technology (TU Delft), Delft, The Netherlands. His research interests include flexible digital computing-in-memory (CIM), low-power and low-latency system-level design optimization for spiking neural network (SNN) accelerators, and design space exploration (DSE) tools.
\end{IEEEbiography}
\vspace{-1.5cm}

\begin{IEEEbiography}[{\includegraphics[width=1in,height=1.25in,clip,keepaspectratio]{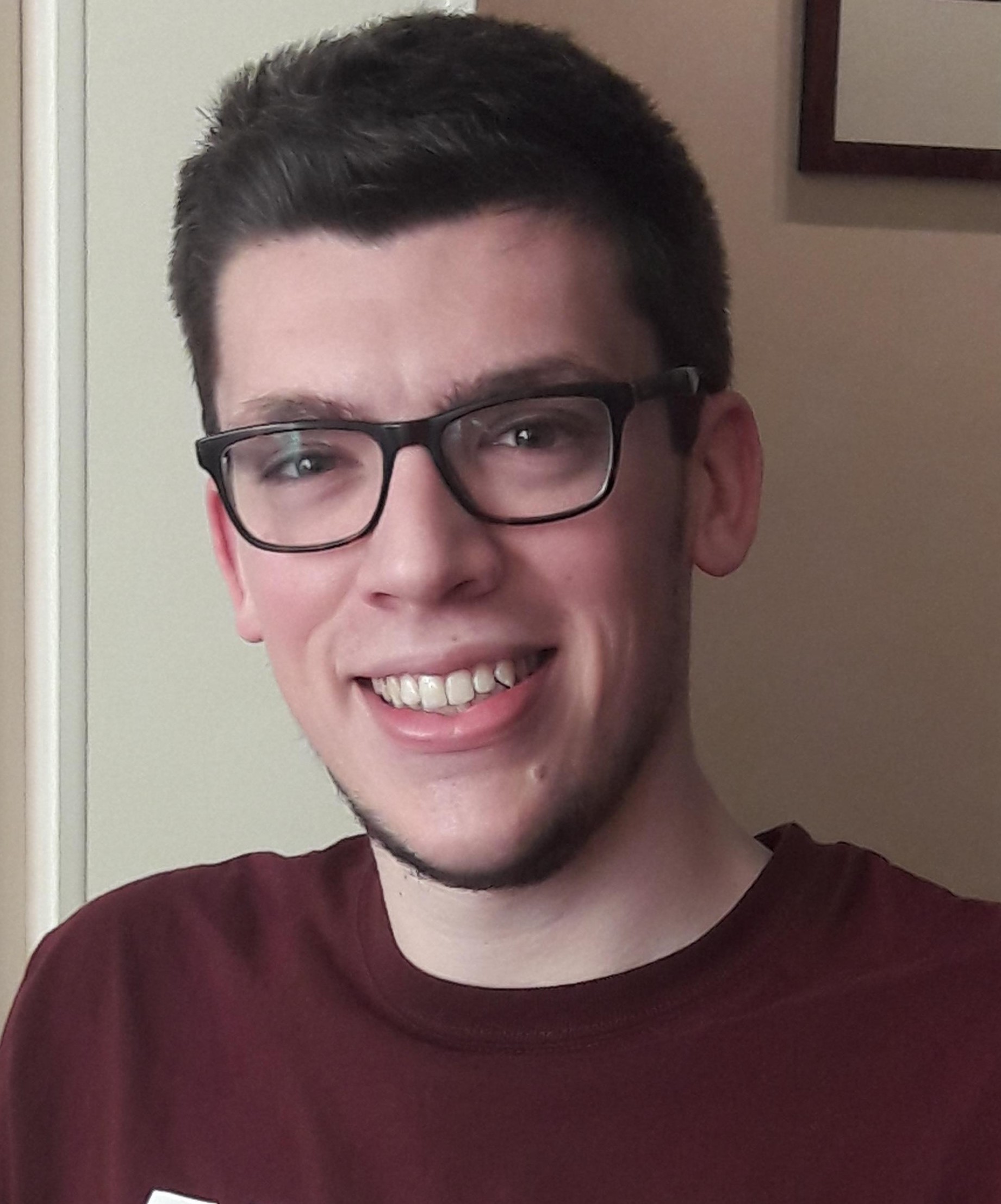}}]{Adrian Kneip}
(Member, IEEE) received the M.Sc. degree (summa cum laude) and Ph.D. degree in electrical engineering from the Universit\'e catholique de Louvain (UCLouvain), in 2019 and 2024 respectively. He is now a postdoctoral research fellow at the Delft University of Technology and at the KU Leuven. His research interest includes the design of mixed-signal processor chips for edge AI, from ultra-low-power to high-performance applications. He has a particular interest for SRAM-based in-memory computing and hardware/software co-design aspects. He is also interested in event-based processing, with a focus on event-driven graph neural networks and their hardware acceleration. Dr. Kneip is the author or co-author of several research papers in IEEE conferences and journals, receiving the Best Student Paper Award for the 2022's ESSCERC conference. He also serves as reviewer for various IEEE SSCS and CAS journals, and was Student Representative to the IEEE Benelux Section from 2020 to 2023.
\end{IEEEbiography}
\vspace{-1.5cm}

\begin{IEEEbiography}[{\includegraphics[width=1.25in,height=1.3in,clip,keepaspectratio]{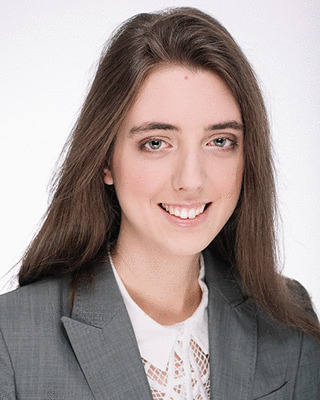}}]{Charlotte Frenkel}
(Senior Member, IEEE) received the M.Sc. degree (\textit{summa cum laude}) in Electromechanical Engineering and the Ph.D. degree in Engineering Science from Universit\'e catholique de Louvain (UCLouvain), Louvain-la-Neuve, Belgium in 2015 and 2020, respectively. In February 2020, she joined the Institute of Neuroinformatics, UZH and ETH Zurich, Switzerland, as a postdoctoral researcher. She is an Assistant Professor at Delft University of Technology, Delft, The Netherlands, since July 2022, and a Research Scientist at Google since February 2026.

Her research aims at bridging the bottom-up (bio-inspired) and top-down (engineering-driven) design approaches toward neuromorphic intelligence, with a focus on hardware-algorithm co-design for (Neuro)AI, digital hardware accelerators, and brain-inspired on-device learning.

Dr. Frenkel received a best paper award at the IEEE International Symposium on Circuits and Systems (ISCAS) 2020 conference in the \textit{Neural Networks} track, and her Ph.D. thesis was awarded the FNRS-FWO / Nokia Bell Scientific Award 2021 and the FNRS-FWO / IBM Innovation Award 2021. In 2023, she was awarded prestigious Veni and AiNed Fellowship grants from the Dutch Research Council (NWO). She presented several invited talks, including keynotes at the tinyML EMEA technical forum 2021 and at the Neuro-Inspired Computational Elements (NICE) neuromorphic conference 2021. She serves or has served as a program co-chair of NICE 2023-2024 and of the tinyML Research Symposium 2024, as a TPC member of IEEE ISSCC for 2027 and IEEE ESSERC for 2022-2024, and as an associate editor for the IEEE Transactions on Biomedical Circuits and Systems for 2022-2025.
\end{IEEEbiography}